\documentclass[journal]{IEEEtran}
\IEEEoverridecommandlockouts

\usepackage{cite}
\usepackage{amsmath,amssymb,amsfonts}
\usepackage{algorithm}
\usepackage[noend]{algorithmic}
\usepackage{graphicx}
\usepackage{subcaption}
\usepackage{textcomp}
\usepackage{xcolor}
\usepackage{caption}
\usepackage{subcaption}
\usepackage{booktabs}
\usepackage{multirow}
\usepackage{float}
\usepackage{stfloats}
\usepackage{enumitem}
\usepackage[normalem]{ulem}
\useunder{\uline}{\ul}{}
\def\BibTeX{{\rm B\kern-.05em{\sc i\kern-.025em b}\kern-.08em
    T\kern-.1667em\lower.7ex\hbox{E}\kern-.125emX}}

\renewcommand{\arraystretch}{0.6}
\begin{document}

\title{HASSLE: A Self-Supervised Learning Enhanced Hijacking Attack on Vertical Federated Learning
}

\author{Weiyang He, Chip-Hong Chang,~\IEEEmembership{Fellow, IEEE}
\thanks{Weiyang He and Chip-Hong Chang are with the School of Electrical and Electronic Engineering, Nanyang Technological University, Singapore}
}

\maketitle

\begin{abstract}
Vertical Federated Learning (VFL) enables an orchestrating active party to perform a machine learning task by cooperating with passive parties that provide additional task-related features for the same training data entities. While prior research has leveraged the privacy vulnerability of VFL to compromise its integrity through a combination of label inference and backdoor attacks, their effectiveness is constrained by the low label inference precision and suboptimal backdoor injection conditions. To facilitate a more rigorous security evaluation on VFL without these limitations, we propose \textit{HASSLE}, a hijacking attack framework composed of a gradient-direction-based label inference module and an adversarial embedding generation algorithm enhanced by self-supervised learning. \textit{HASSLE} accurately identifies private samples associated with a targeted label using only a single known instance of that label. In the two-party scenario, it demonstrates strong performance with an attack success rate (ASR) of over 99\% across four datasets, including both image and tabular modalities, and achieves 85\% ASR on the more complex CIFAR-100 dataset. Evaluation of \textit{HASSLE} against 8 potential defenses further highlights its significant threat while providing new insights into building a trustworthy VFL system.
\end{abstract}

\begin{IEEEkeywords}
Federated Learning, Label-inference Attack, Backdoor Attack, Adversarial Examples
\end{IEEEkeywords}

\section{Introduction}
\IEEEPARstart{F}{ueled} by the recent success of machine learning models trained from a large scale of data, increasing hope has been put on Federated Learning (FL) to further enhance model performance without violating numerous privacy regulations \cite{gdpr, ccpa}. FL enables multiple parties to train machine learning models without exchanging private data samples.  
Depending on how the data is partitioned, it can be categorized into horizontal FL (HFL) and vertical FL (VFL). In HFL, the partition is based on individual data sample, whereas in VFL, the division is based on features of a shared identity. An intuitive example of VFL involves an insurance company aiming to estimate the likelihood of successfully selling its products to new clients. Instead of training a prediction model using its own features like clients' health information and past purchase record, the insurance company can cooperate with a bank that serves the same group of clients. By leveraging additional client features, such as their loan and credit information, the two parties can train a more accurate model. In this example, the insurance company acts as the active party who initiates the VFL and owns the task labels, while the bank serves as a passive party, contributing supplementary data. VFL holds immense potential and  is gradually gaining ground in privacy-sensitive sectors like healthcare and finance \cite{long2020federated, cha2021implementing, song2021federated}.

Though promising, the security and privacy vulnerability of VFL is not well studied, especially when its variant HFL has been found susceptible to various poisoning and privacy leakage attacks\cite{rodriguez2023survey}. In a dependable VFL system, no participating party should learn any private information about other parties and the model prediction result should be sensible based on reliable training and test data from all parties. Nevertheless, preliminary research has revealed that a passive party can confidently infer the private labels of its training samples \cite{fu2022label}, and hijack the final model prediction by sending falsified embeddings to the active party \cite{pang2022adi}. Even more concerning is the leaked label information can pave the way for an even more vicious backdoor attack\cite{naseri2024badvfl} in VFL. The attacker can utilize the extracted label information to selectively poison its training samples. Through the implanted backdoor trigger, the attacker can inconspicuously hijack the VFL inference for targeted misclassification. 

Although previously proposed hijacking attacks claim to achieve a high attack success rate, our evaluation reveals that their effectiveness is limited in practice due to two main issues. Firstly, the label inference module often fails by incorrectly identifying poisoned samples with non-target labels, which reduces the overall efficacy of the attack. Secondly, the subsequent poisoning injection at the later stage of training establishes a weak connection between target label and trigger due to a low poison injection rate and the limited feature saliency held by the attacker.

\begin{figure}[t]
    \includegraphics[width=\linewidth]{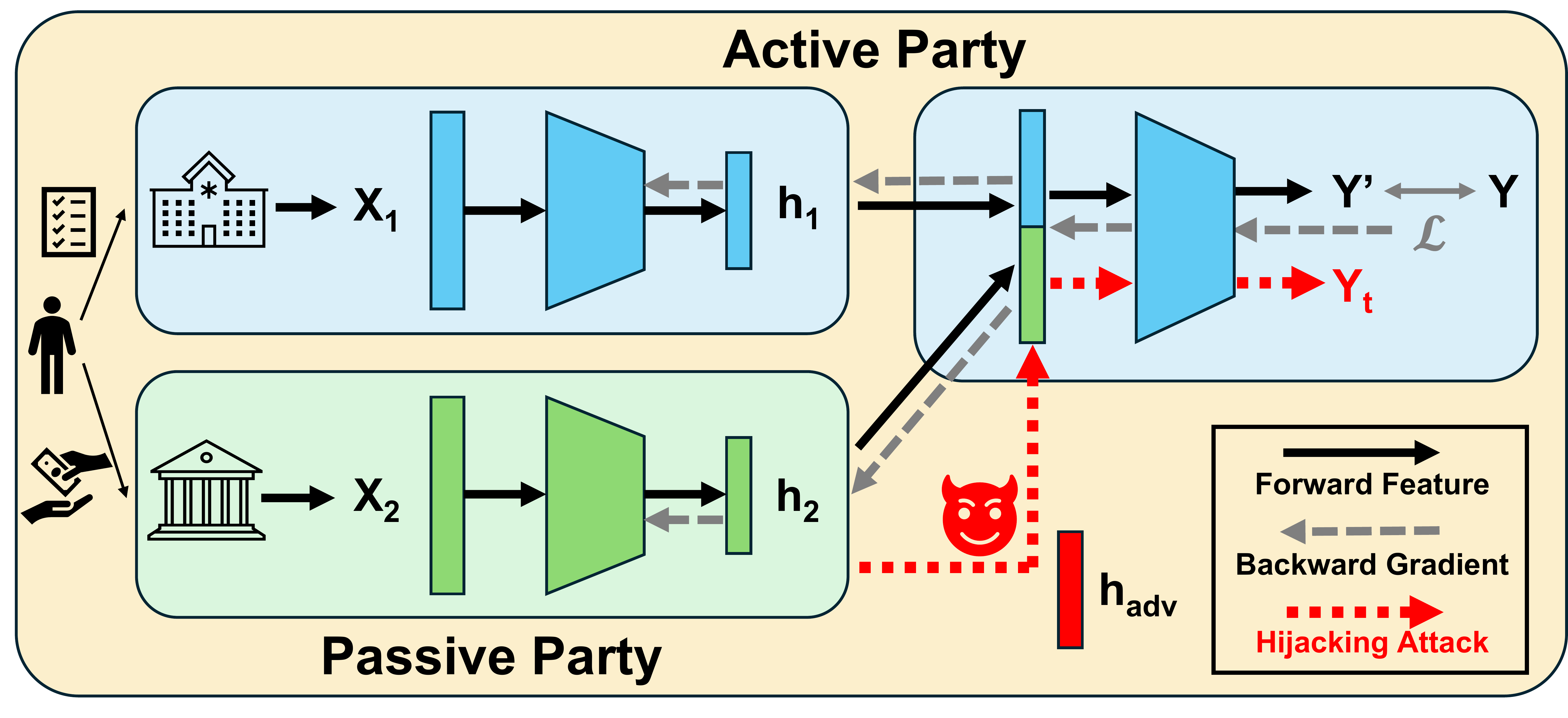}
    \caption{The forward/backward process of VFL framework and the hijacking attack.}
    \label{splitvfl}
\end{figure}

To bridge the gap, we propose a novel VFL attack framework called \textbf{HASSLE} (\underline{H}ijacking \underline{A}ttack on VFL model with \underline{S}elf-\underline{S}upervised \underline{L}earning \underline{E}nhancement). Our major contributions in HASSLE are summarized as follows: 
\begin{itemize}[leftmargin=*]
    \item We formally analyze the causation between the direction of returned gradient and the corresponding sample label in VFL, and design a robust and stealthy label inference attack that achieves high precision with only one known sample. 
    \item We design a new adversarial embedding method that can force the model to predict an attacker-designated label. The adversarial embedding replaces the original embedding of target-label samples during each round of VFL training. The returned gradients are, in turn, used to iteratively optimize the adversarial embedding. The incorporation of adversarial examples into poisoning attack significantly enhances the success rate of our hijacking attack. 
    \item We draw inspiration from representation learning that an encoder capable of extracting better representation from raw data can provide better separability for the downstream classifier. To gain greater feature saliency during training, we initialize the attacker's bottom model using self-supervised learning on the attacker's own dataset to strengthen its influence on the VFL model. 
    \item We evaluate HASSLE on five datasets. In two-party VFL setting, its label inference and hijacking attacks achieve 100\% precision and over 85\% attack success rate, respectively. We also perform additional sensitivity analysis on HASSLE with different VFL hyperparameter settings to demonstrate its attack performance consistency. The robustness of HASSLE is also attested by 8 potential defense methods. The findings provide actionable insights into improving the VFL security.

\end{itemize} 

\section{Preliminaries}

\subsection{Vertical Federated Learning}
\begin{figure}[!h]
    \includegraphics[width=\linewidth]{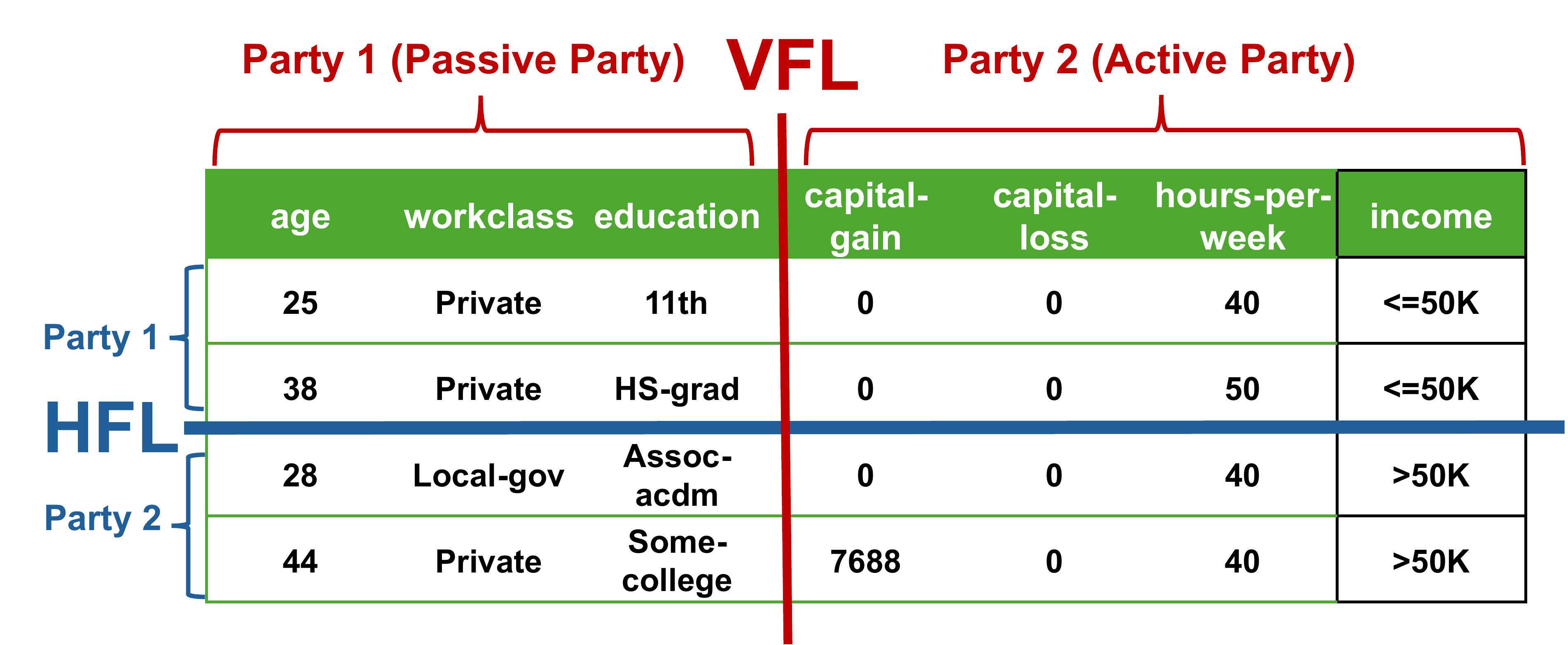}
    \caption{Different partitions between HFL and VFL. Income is the label to be predicted.}
    \label{vflhfl}
\end{figure}

Unlike HFL, where all participants own unique data samples and share their locally trained model during training, VFL is employed when different parties hold a limited and disjoint set of features of the same entity. To better understand their difference, one can think of it as slicing the table in Fig.~\ref{vflhfl} either horizontally or vertically to allocate each of these segments to different parties. The orchestrator who uniquely owns the label of the learning task in VFL is called the active party, while others are the passive parties \cite{liu2024vertical} . 

SplitVFL is a widely adopted framework in VFL due to its ability to achieve high accuracy and flexibility in accommodating different models for each participating party. During the training phase, all parties first encode their feature subsets locally using their respective bottom models and then send the resulting embeddings to the active party. The active party then sends the calculated loss gradient back to all parties to facilitate backward propagation and local model parameter updates. During the inference phase, each party processes its corresponding subset of features and sends embeddings to the active party for fusion and final inference. This paper applies SplitVFL to classification tasks, as it represents the most common use case in practice.

\subsection{Privacy and Security Threats in VFL}
The privacy and security flaws in HFL have been well studied in recent years \cite{rodriguez2023survey}. Privacy leakage arises from the strong correlation between the model updates and private data of participating parties. Security threats, on the other hand, stem from the inability of the orchestrating party to validate the integrity of raw training data. Although attack techniques designed for HFL are not directly transferable to VFL, the privacy-preserving capability and intermediate data exchange protocol in VFL signify that it can be susceptible to similar threats. Existing attacks on VFL are reviewed as follows.

\subsubsection{Privacy Attack}
A key requirement of FL is to preserve the privacy of each participating party such that no party should learn about the private information from other parties. However, without proper defense, a malicious passive party can infer the private label held by the active party. Contrary to intuition, intermediate gradients returned by the active party embed sufficient information about the labels used in the loss calculation. In extreme cases, such as when the top model only consists of a summation and softmax operation, the attacker can identify the ground truth label by simply finding the index of the returned gradient with a negative value \cite{fu2022label,zou2022defending}. With a more complex top model like a fully-connected neural network with multiple layers, Li et al. \cite{li2022label} demonstrated that gradients calculated with the same label tend to have similar direction and norm in a binary classification task. Building on this intuition, BadVFL \cite{xuan2023practical} and VILLAIN \cite{bai2023villain} attacks replace some uploading embeddings with one that is known to belong to the target label. Based on the cosine similarity or norm ratio between their returned gradients that are affected by the replacement and those that are not, they determine if these replaced embeddings also belong to the target label. Sun et al. \cite{sun2022label} established the connection between forward embeddings to be sent from the attacker and their corresponding labels by conducting a spectral clustering on the embeddings. By leveraging the learned knowledge in the bottom model, Fu et al. \cite{fu2022label} trained a surrogate top model with an auxiliary labeled dataset to achieve high accuracy on its training samples with partial features. However, these attacks face practical limitations due to additional assumptions about the system being attacked, such as binary classification tasks, single-layer top models, or the availability of a sizable auxiliary dataset.

\subsubsection{Hijacking Attack}
Traditional machine learning is susceptible to poisoning attacks and evasion attacks, both aim at maliciously altering the model behaviour at inference time. In contrast, VFL involves collaboration among multiple parties to produce the model output, which requires an attacker to amplify its own influence to manipulate the output. Therefore, we categorize all such attacks as \textit{Hijacking Attack} in VFL. As it is trivial for an active party to change the model output, the attackers are assumed to be some of the passive parties. Hijacking attack can be implemented in a form similar to adversarial examples \cite{goodfellow2014explaining}, where the attacker feeds a perturbed feature subset that would maximize the top model's output probability for its target label \cite{pang2022adi}. Alternatively, an attacker can poison the top model through backdoor attack to hijack the VFL model. Through mean-shift algorithm, He et al. \cite{he2023backdoor} replaces a set of training embeddings belonging to a specific target label with a mode in a dense region surrounded by target-label embeddings. BadVFL \cite{xuan2023practical} adds trigger to a portion of its training samples that are inferred to be the target label. The top model will output the target label when the attacker feeds it with samples that contain the trigger in test time. VILLAIN \cite{bai2023villain} increases the attack's effectiveness by adding the trigger in the submitted embedding instead of the input domain.  Leveraging the surrogate model of \cite{fu2022label}, Gu et al. \cite{gu2023lr} found strongly target-label correlated adversarial embeddings through model inversion. Qiu et al. \cite{qiu2024hijack} achieves the same goal with a similar approach. With the adversarial embedding, the trigger pattern in the raw feature domain is learned by training the bottom model with poisoned samples. To succeed in these attacks, knowledge of the ground truth label of training samples is needed, which is inaccessible to any passive parties. Any wrong label predictions by the attacker can result in significant degradation of attack effectiveness since the trigger may be falsely trained to link with a non-target label. Moreover, the acquisition of label information by many previously proposed label inference attacks can only be completed during the later stage of training, which delays the backdoor injection and reduces their efficacy.

\section{Proposed Methodology}

\subsection{Formulation of VFL Framework}
The VFL framework aims to solve a classification task with $C$ possible classes and $n$ training samples, where $K$ parties collaborate to infer the task label $y_i$ of a sample with ID $i$ using the partial features $x_{i,k}$ owned by the respective $k^{th}$ parties. This objective is formulated as follows:
    \[\mathcal{L}=\mathbb{E}_{i\sim U(1,n)} [\emph{CE}(g(f_1(x_{i,1})|| f_2(x_{i,2})||...||f_K(x_{i,K})); y_i)]\tag{1}\label{eq:1}\]
where $U(1,n)$ denotes a discrete uniform distribution with integer values from $1$ to $n$,  $g$ is the top model of the active party, and $f_i$ is the bottom model of the $i^{th}$ party. The cross entropy loss $\emph{CE}$ is used for optimizing the classification task.

The intermediate embedding calculated by the $k^{th}$ party from the $i^{th}$ sample is denoted as $h_{i,k} = f_k(x_{i,k})$. The top model $g$ (typically a shallow neural network) takes a vector $h_i$ concatenated from $\{h_{i,k}\}_{k=1}^K$ as input and produces a probability score vector $s_i \in \mathbb{R}^C$, which is the output of the softmax function of logits $l_i$.

During the backward process, the active party calculates $dh_i$, the gradient of the loss $\mathcal{L}$ with respect to $h_i$ by:
\[dh_i = \frac{\partial \mathcal{L}}{\partial h_i} = \frac{\partial \mathcal{L}}{\partial s_i} \frac{\partial s_i}{\partial l_i} \frac{\partial l_i}{\partial h_i}\tag{2}\label{eq:2}\]
Then the gradient $dh_i$ is sliced into $K$ vectors, $\{dh_{i,k}\}_{k=1}^K$, and sent to the $K$ corresponding parties.

\subsection{Attack Model}

\subsubsection{Adversary's Goal}
The attacker's objective is to force the model to predict an adversary-designated target label when an attack embedding is submitted. The classification accuracy of the model should not be impaired when a normal embedding is submitted. Furthermore, the adversary's action should be inconspicuous to avoid being detected by the active party.

\subsubsection{Adversary's Knowledge and Capability}
The adversary is assumed to be a participating passive party who knows the total number of task labels. Additionally, he has the partial features and IDs of the training samples, but not their task labels. Moreover, we hypothesize that the adversary can obtain the ID of one training sample with the desired target label through unauthorized access with sniffed password or stolen credentials, phishing, exploiting software weaknesses, installing rootkit, colluding with a data provider, etc. In VFL systems, individual data providers often serve as common customers for both the adversary and the active party, making them potential targets for collusion driven by financial incentives. During the training stage, the adversary can receive the returned gradients from the active party but has zero knowledge about the data features owned by other parties, their submitted embeddings or their received gradients. Furthermore, the adversary is agnostic to the top model parameters and architectures.

Since the partial features and bottom models are privately held by each party, the adversary has full control over the training of his bottom model, and can submit any embeddings to the active party during both training and inference phases. 

\begin{figure}
    \centering
    \includegraphics[width=0.75\linewidth]{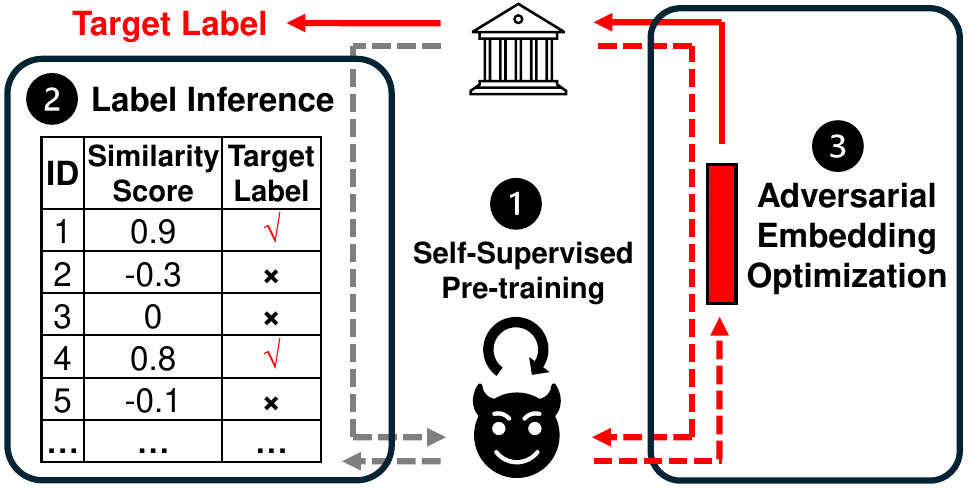}
    \caption{General framework of HASSLE.}
\end{figure}

\subsection{Improved Gradient-Direction-Based Label Inference Attack} \label{3-c}
The aim of label inference attack (LIA) is to obtain the necessary label information to either poison the target-label data or introduce targeted perturbation for the hijacking attack on VFL. To meet the more stringent attack model requirements and improve the attack success rate (ASR) of our hijacking attack, the LIA must fulfill all the following desirable properties: 

\begin{enumerate}
    \item It requires minimal auxiliary label information.
    \item It can achieve high precision on predicted sample.
    \item It operates effectively in early training epochs.
    \item It is stealthy and highly evasive.
\end{enumerate}

Firstly, the attacker should not be assumed to possess sizable ground truth labels of its training samples, as such private labels owned by the active party are usually well-protected and is costly to obtain even a small portion of them. The attacker can only acquire the index of a sample belonging to the target class, which is the minimum amount of information required to perform an LIA in a multi-class classification task. Secondly, the ASR of most hijacking attacks are highly affected by the LIA accuracy, since a falsely predicted non-target-class sample can either weaken the model's attention on backdoor trigger or direct the hijacking embedding away from the target decision boundary. Thirdly, backdoor trigger injections should start as early as possible in the training epochs to allow the top model to more easily learn the association of the trigger and target label, leading to a stronger and more persistent backdoor. Finally, abnormal embeddings that can be easily detected or affect the main task accuracy should not be sent to the active party.
In \cite{li2022label}, it is analytically shown that two returned gradients have positive cosine similarity if their corresponding task labels are the same in a binary classification task. This finding is extended to the multi-class setting using an embedding-swapping technique in \cite{xuan2023practical}. However, these two LIA schemes satisfy only Properties 1 and 3. Despite achieving convincing accuracy in binary classification task, they deteriorate significantly when the top model consists of more than one layer in a multi-class scenario. To overcome this severe limitation, we perform a formal analysis on the gradients received by the adversary to identify its root cause.

Suppose the top model is a single-layer network formulated as $l_i=wh_i$, where $w\in \mathbb{R}^{C\times H}$, (\ref{eq:2}) can be rewritten as:

\vspace{-0.2cm}

\[
dh_i = w^\top\frac{\partial \mathcal{L}}{\partial l_i} = w^\top(s_i-y_i) = w^\top\epsilon_i \tag{3} 
\label{eq:3}
\]

\vspace{-0.2cm}
where   $\top$ denotes the matrix transpose, $s_i$ is the model predicted categorical distribution, and $y_i$ is the one-hot vector representing the ground truth label. 

It is trivial to show that if only one element in the error vector $\epsilon_i$ is negative, its index is the ground truth class. The gradient $dh_i$ can thus be constructed as a linear combination of the column vectors of $w^\top$, with one of these vectors connected to the ground truth logit having the only negative coefficient, as illustrated below:

\vspace{-0.6cm}

\begingroup
\renewcommand{\arraystretch}{0.5}
\begin{align}
    \text{$H$}\left\{
\begin{bmatrix}
    \\ 
    dh_i \\
    \\
\end{bmatrix}\right.
&=
\overbrace{
\begin{bmatrix}
   &  &  \\
   & w^\top &  \\
   &  & 
\end{bmatrix} 
}^{C}
\begin{bmatrix}
    \\ 
   \epsilon_i \\
   \\
\end{bmatrix} \notag
\\
&=
\begin{bmatrix}
     \\
   w_1 \dots w_{y_i} \dots w_C\\
    \\
 \end{bmatrix} 
 \begin{bmatrix}
    + \\
    \vdots\\ 
    - \\
    \vdots\\
    + \\
 \end{bmatrix}
 \tag{4}
\label{eq:4}
\end{align}
\endgroup

\vspace{-0.5cm}

For a binary classification,

\vspace{-0.3cm}

\[dh_i = \epsilon_{i,1} w_1 + \epsilon_{i,2} w_2 = \epsilon_{i,1} (w_1 - w_2) \tag{5}\label{eq:5}\]

\noindent where $\epsilon_{i,c}$ and $w_c$ represent the $c^{th}$ index of $\epsilon$ and the weight vector connected to the $c^{th}$ logit, respectively. Eq. (\ref{eq:5}) agrees with the observation of \cite{li2022label} that the gradients of two samples have positive cosine similarity if they have the same label, and negative cosine similarity otherwise, given the negligible change in their weight vectors within one training epoch. 

For multi-class classification, the directional similarity cannot be directly derived. However, it is reasonable to approximate the categorical distribution produced by the randomly initialized model in the early stage of VFL training as a uniform distribution given any training samples. Then, by substituting $s_{i,c}=\frac{1}{C}$ into (\ref{eq:3}), we have

\vspace{-0.4cm}

\begin{align}
    dh_i &= \frac{1}{C}w_1 + \dots + \left(\frac{1}{C} - 1\right)w_{y_i} + \dots + \frac{1}{C}w_C \notag
    \\
         &=  -\left(1 - \frac{1}{C}\right)w_{y_i} + \frac{1}{C}w_1 + \dots + \frac{1}{C}w_C \notag
         \tag{6} \label{eq:6}
\end{align}

Since the number of task labels $C$ is at least 3, the magnitudes of the weight vector coefficients corresponding to the ground truth label substantially outweigh those of the other labels. Moreover, the directions of these weight vectors are expected to be uniformly distributed in early training, which results in the direction of gradient $dh_i$ dominated by $w_{y_i}$, with minimal alignment from the remaining vectors. This enables an LIA based on the cosine similarity between a gradient associated with a known label and one associated with an unknown label in the first training epoch.

If the top model uses multiple layers with the Rectified Linear Unit (ReLU) activation function, then the LIA based on gradient direction during early training will be impaired and (\ref{eq:6}) is no longer valid. For simplicity and without loss of generality, we assume that the top model has two layers, which is formulated as $l_i = w^2\sigma(w^1h_i)$, where $w^l$ refers to the model weight of the $l^{th}$ layer and $\sigma$ the ReLU activation function. Then, (\ref{eq:2}) can be re-expressed as

\vspace{-0.3cm}

\[dh_i = {w^1}^\top\sigma'(w^1h_i){w^2}^\top(s_i-y_i) \tag{7} \label{eq:7}\]

\vspace{-0.3cm}

Likewise, we decompose (\ref{eq:7}) into

\vspace{-0.5cm}

\begingroup
\renewcommand{\arraystretch}{0.6}
\begin{align}
    H \left\{
    \begin{bmatrix}
        \\
        dh_i \\
        \\
    \end{bmatrix}\right.
    &=
    \overbrace{
    \begin{bmatrix}
        &  &  \\
        & {w^1}^\top &  \\
        &  & 
    \end{bmatrix}
    }^{M}
    \overbrace{ 
    \begin{bmatrix}
        &  &  \\
        & \sigma' &  \\
        &  & 
    \end{bmatrix}
    }^{M} 
    \overbrace{
    \begin{bmatrix}
        &  &  \\
        & {w^2}^\top &  \\
        &  & 
    \end{bmatrix} 
    }^{C}
    \begin{bmatrix}
        \\ 
        \epsilon_i \\
        \\
    \end{bmatrix} \notag \\
    &\hspace{-1.5cm}=
    {w^1}^\top
    \operatorname{diag}(\sigma(w^1h_i))
    \begin{bmatrix}
        \\
        w_1^2 \dots w_{y_i}^2 \dots w_C^2 \\
        \\
    \end{bmatrix} 
    \begin{bmatrix}
        + \\
        \vdots\\ 
        - \\
        \vdots\\
        + \\
    \end{bmatrix}
    \tag{8}
\label{eq:8}
\end{align}
\endgroup

\vspace{-0.2cm}

\noindent where $diag(\cdot)$ denotes a diagonal matrix constructed from a vector and $M$ is the dimension of the first neuron layer of the top model. 

Comparing with (\ref{eq:4}) for the single-layer, two additional terms, ${w^1}^\top$ and $\sigma'$, have been introduced into the expression of $dh_i$ in (\ref{eq:8}). While the term ${w^2}^\top\epsilon_i$ maintains the correlation between the direction and ground truth label as explicated in the single-layer setting, the new term ${w_1}^\top \operatorname{diag}(\sigma(w^1h_i))$ introduces noise that obscures this correlation. Specifically, $\operatorname{diag}(\sigma(w^1h_i))$ filters out the dimensions in ${w^2}^\top\epsilon_i$, where $w^1h_i\leq0$, and ${w^1}^\top$ applies a linear transformation to project the vector from the hidden layer dimension to that of the embedding space. In early training stage, since $w_1$ is randomly distributed and $\sigma(w^1h_i)$ is largely independent of the label, this term can dramatically diminish the directional similarity between pairs of $dh_i$ with the same label. Moreover, as the number of layers in the top model increases, the directional clue becomes weaker due to the repeated application of linear transformation and ReLU filtering.

\begin{figure}[h]
    \hspace{-0.3cm}
    \begin{subfigure}{0.48\linewidth}
        \includegraphics[height=3.2cm]{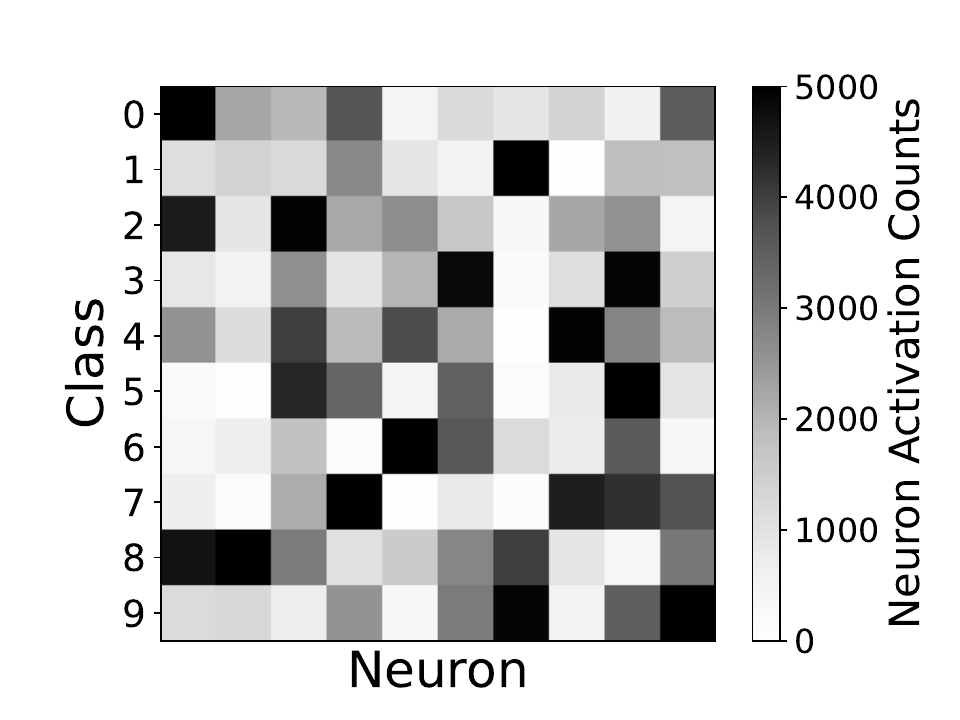}
        \caption{}
        \label{heatmap}
    \end{subfigure}
    \begin{subfigure}{0.51\linewidth}
        \includegraphics[height=3.2cm]{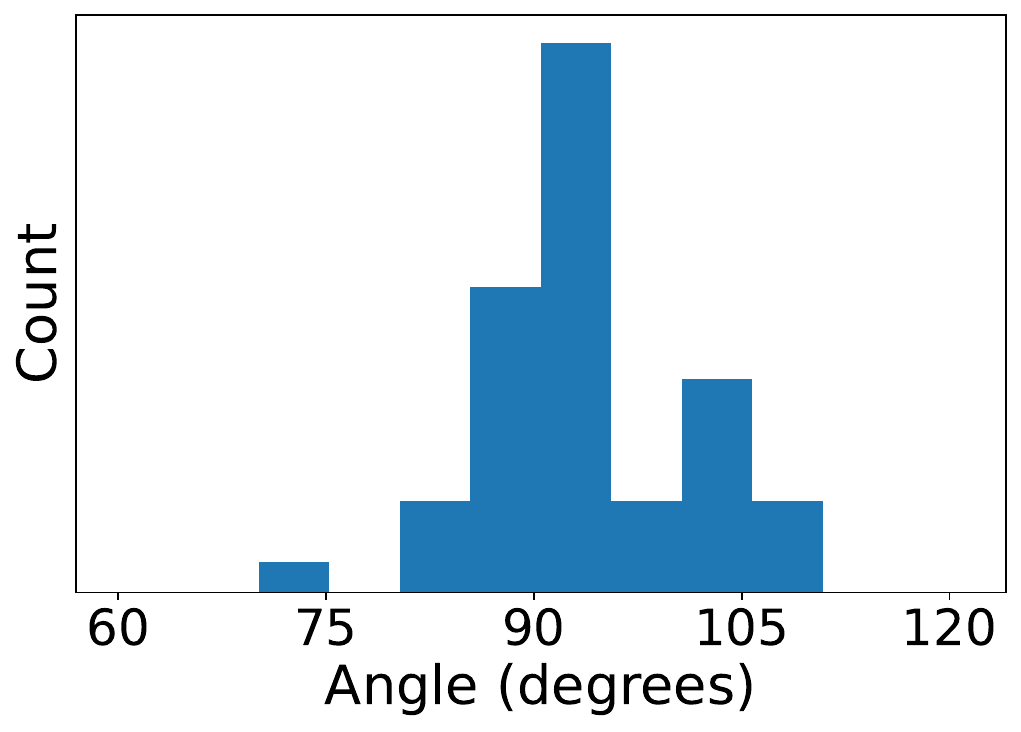}
        \caption{}
        \label{angle}
    \end{subfigure}
    \caption{(a) Heatmap of neuron activation counts at the first ReLU layer of the top model using CIFAR-10 training dataset. Each class of data has distinct frequently activated neurons, which helps to maintain the correlation between the ground truth label and the returned gradient direction. (b) Distribution of angles between the column vectors in ${w^1}^\top$ at one-third of the total training epochs on CIFAR-10 dataset.}
\end{figure}

\textit{Is there a way to preserve the direction-label correlation with the linear transformation and ReLU filtering terms?} Surprisingly, a VFL model with well-trained bottom and top models help shape this property to a certain degree. In a typical SplitVFL framework, the bottom models extract the low-level features from the raw data samples and the top model makes prediction based on the high-level features submitted by all parties. Generally, the training samples with the same label have a similar set of high-level features, and share similar hidden layer neuron activations. By contrast, there exists a distinct difference between the activations of samples from different classes. This characteristic is visualized in Fig. \ref{heatmap}, which allows the correlation between gradient direction and label to pass through the diagonal matrix of $\sigma'$ with negligible perturbation. Similar idea can be applied to the additional weight matrix ${w^1}^\top$. To begin with, we rewrite (\ref{eq:7}) by combining the last three terms into $\frac{\partial \mathcal{L}}{\partial u}$:

\vspace{-0.3cm}

\begingroup
\renewcommand{\arraystretch}{0.6}
\begin{equation}
    dh_i =
    {w^1}^\top \frac{\partial \mathcal{L}}{\partial u}
    = 
    \begin{bmatrix}
        \\
        w_1^1 \dots w_M^1 \\
        \\
    \end{bmatrix}
    \begin{bmatrix}
          \\
         \frac{\partial \mathcal{L}}{\partial u}  \\
          \\ 
    \end{bmatrix}
    \tag{9}
    \label{eq:9}
\end{equation}
\endgroup

\vspace{-0.3cm}

The individual column vectors $\{w_m^1\}_{m=1}^M$ in ${w^1}^\top$ describe how different features in the next layer are composed from the features in the previous layer. The directions of these column vectors are assumed to be highly divergent in a modestly trained network because different high-level features should depend on different low-level features if they are meaningful. Fig. \ref{angle} indeed demonstrates that the column vectors $\{w_m^1\}_{m=1}^M$ are approximately orthogonal to one another. By viewing $dh_i$ as a linear combination of these column vectors using coefficients from $\frac{\partial \mathcal{L}}{\partial u}$, we can conclude that the gradient is correlated with the ground truth label.

However, in a modestly trained model, (\ref{eq:6}) is no longer valid and this affects the directional clues about the label in $\frac{\partial \mathcal{L}}{\partial u}$. To overcome this, we will explore a different property from $\epsilon_i$: 

\vspace{-0.3cm}

\[\epsilon_{i,y_i} = - \sum_{c \in \{1, \dots, C\} \setminus \{y_i\}} \epsilon_{i, c} \tag{10} \label{eq:10}\]

\vspace{-0.1cm}

To guarantee that the ground truth label linked weight vector $w_{y_i}^2$ will  dominate the direction of ${w^2}^\top\epsilon_i$ as in (\ref{eq:6}), the largest element in $\{\epsilon_{i,c}\}_{c \in \{1, \dots, C\} \setminus \{y_i\}}$ should not be close to the opposite number of $\epsilon_{i, y_i}$. This condition is satisfied when the second largest probability score is not multiple times larger than the third largest probability score. Under such condition, ${w^2}^\top(s_i-y_i)$ will retain the label-related directional information as it propagates to $dh_i$.

We now elucidate our improved gradient-direction based LIA. Due to the direction disruption from the linear projection and ReLU filtering caused by the random initialization, the received gradient information in the first epoch is skipped. Starting from the second training epoch until an adversary-selected epoch $E_a$, the cosine similarity between the gradient of the known sample with the target label and all the other gradients are calculated for every epoch. These similarity scores are recorded for $E_a-1$ epochs and averaged for a more robust attack. After epoch $E_a$, the attacker will have a similarity score for each training sample. Since the attacker has the knowledge of the number of task labels and the number of total samples, he can estimate the number of samples with the target label $n_t$ by a simple division. Finally, the samples with the target label can be determined by selecting the IDs of the $n_t$ samples with the highest similarity scores. To improve the LIA accuracy and account for cases where $n_t$ may be lower than expected, the adversary can set a tunable parameter $r$ to only sift out the top $\frac{n_t}{r}$ samples. Our LIA is simple and practical. It extends the applicable scenarios from \cite{li2022label} and alleviates the detectability of the blatant embedding swapping technique~\cite{xuan2023practical} that replaces huge fraction of submitted embeddings with a constant embedding.

\subsection{HASSLE Hijacking Attack}
After revealing a decent amount of target-label sample IDs, HASSLE hijacking attack is launched with the adversarial embedding optimization algorithm and self-supervised learning enhancement technique. 

\subsubsection{Adversarial Embedding Optimization}
The objective of our hijacking attack is formulated as follows:

\vspace{-0.3cm}

\[\min \mathbb{E}_{h_1, \dots, h_K} [\emph{CE}(g(h_1||h_{adv}||...||h_K); y_t)]\tag{11}\label{eq:11}\]

\vspace{-0.1cm}

\noindent where $h_k$ represents the forward embedding submitted by the $k^{th}$ party and $h_{adv}$ corresponds to the attacker's adversarial embedding. To achieve a high ASR, this adversarial embedding should force the top model $g$ to predict the target label regardless of the embeddings submitted by other parties.

Previous works realize this by two different attacks: poisoning that updates the top model parameter $g$ and adversarial perturbation that updates the adversarial embedding $h_{adv}$. However, we discover that these two attacks can be naturally combined in the context of VFL. This is because the returned gradient $dh_{adv}$ received by the attacker provides a meaningful update direction for the submitted embedding to steer it towards the decision region associated with the supervising label. If the supervising label matches the attacker's target label, the gradient $dh_{adv}$ can be directly used to optimize $h_{adv}$. The prerequisite of obtaining $dh_{adv}$ towards the target label coincides with the necessity to replace the original embedding with a poisoned embedding. Thus, the adversarial embedding can simultaneously be used as a poisoned sample associated with the target label that constantly appears in the training.

\begin{algorithm}[t]
    \renewcommand{\algorithmicrequire}{\textbf{Input:}}
    \renewcommand{\algorithmicensure}{\textbf{Output:}}
    \caption{\textit{HASSLE} Attack Framework}
    \label{alg:epoch_specific_training}
    \begin{algorithmic}[1]
    \REQUIRE Training data $D$, total epochs $E$, model parameters $\theta$, known ID $I_k$, ratio $r$, number of training samples $n$
    \ENSURE Trained model parameters $\theta$, optimized adversarial embedding $h_{adv}$

    \STATE $\theta \leftarrow \textsc{Ramdom\_Init()}$
    \STATE $\theta \leftarrow \textsc{SSL\_Pretrain}(\theta,D)$
    \FOR{$e = 1$ to $E$}
        \IF{$2 \leq e \leq E_a$}
            \STATE $I_t \leftarrow \textsc{LIA}(I_k, r)$ \quad /* Infer target sample IDs */
        \ELSIF {$e > E_a$} 
            \FOR{$i = 1$ to $n$}
                \IF{$i \in I_t$}
                    \STATE Submit $h_{adv}$ instead of $h_i$
                    \STATE $h_{adv} \leftarrow h_{adv} - \overline{dh_{adv}}$
                    \STATE $h_{adv} \leftarrow \textsc{Clip}(h_{adv}, \overline{\|h\|})$
                \ENDIF
            \ENDFOR
        \ENDIF
        \STATE $\theta \leftarrow \theta - \eta \nabla \theta$ \quad /* Update bottom model parameters */
    \ENDFOR
    \RETURN $\theta$, $h_{adv}$
    \end{algorithmic}
\end{algorithm}

Based on the above analysis, we first initialize $h_{adv}$ as the embedding of a target-label sample. In subsequent training rounds, we replace the embeddings of those samples that have been found to belong to the target label in our preceding LIA with $h_{adv}$. After receiving the corresponding gradients to the modified embeddings $dh_{adv}$, we update $h_{adv}$ by subtracting the average of $dh_{adv}$ within one batch. To prevent the adversarial embedding from having an overly inflated norm that can be easily detected, we scale it down after each update to match the average norm of all training embeddings. The new $h_{adv}$ is used for embedding substitution in the next training batch and this process is iterated until the VFL training completes. At the end of the training, we obtain an adversarial embedding $h_{adv}$ that effectively hijacks the model to predict a target label. 

Our attack is stealthy because the replaced $h_{adv}$ evolves dynamically and appears differently in every batch. Even within the same batch, the probability of $h_{adv}$ appearing multiple times is low, since the poisoning rate can be as low as 2.5\%. However, a limitation of this method is that the gradient $dh_{adv}$ cannot approximate scenarios where all other submitted embeddings originate from non-target-label samples. As a result, the attack still relies mainly on the poisoning effect to achieve a strong performance.

\subsubsection{Self-Supervised Learning Based Model Initialization}
The SplitVFL paradigm involves a collaborative inference stage where the top model considers the submitted embeddings from all parties sensibly to make a prediction. Therefore, to improve the attack effectiveness, the adversary can make his embeddings contribute more towards the final prediction. Previous works \cite{fu2022label,xuan2023practical,bai2023villain} tried to achieve this goal by increasing the learning rate of the adversary's bottom model. However, this may introduce instability during training and the bottom model may fail to learn high-quality representations, ultimately jeopardizing the final model accuracy. To address this, we turn to the abundant unlabeled samples owned by the adversary. We conjecture that the top model attributes higher importance to the party who learns a better representation at its bottom model, since it provides more useful information related to the task label. Hence, we propose to leverage self-supervised learning (SSL) to pre-train a bottom model using the unlabeled samples before the VFL training begins. It is expected that the top model will assign greater weight towards the adversary's submitted embeddings compared to other parties over the course of the entire training process given their better separability from the start.

\begin{table*}[ht!]
    \caption{Dataset and VFL Training Details}
    \begin{tabular}{@{}c|cccccccccc@{}}
        \toprule
        Dataset    & \begin{tabular}[c]{@{}c@{}}Training \\ Samples\end{tabular} & \begin{tabular}[c]{@{}c@{}}Test \\ Samples\end{tabular} & \begin{tabular}[c]{@{}c@{}}Feature \\ Size\end{tabular} & \begin{tabular}[c]{@{}c@{}}Embedding \\ Dimension\end{tabular} & \begin{tabular}[c]{@{}c@{}}Number of \\ Classes\end{tabular} & \begin{tabular}[c]{@{}c@{}}Number of \\ Top Model Layers\end{tabular} & Batch Size & \begin{tabular}[c]{@{}c@{}}Learning \\ Rate\end{tabular} & Epoch & Metric   \\ \midrule
        CIFAR-10   & 50000                                                       & 10000                                                   & 32$\times$32$\times$3                                                 & 10                                                             & 10                                                           & 3                                                                     & 32         & 0.05                                                     & 20    & Accuracy \\
        CIFAR-100  & 50000                                                       & 10000                                                   & 32$\times$32$\times$3                                                 & 100                                                            & 100                                                          & 3                                                                     & 32         & 0.05                                                     & 40    & Accuracy \\
        ImageNette & 9469                                                        & 3925                                                    & 128$\times$128$\times$3                                               & 10                                                             & 10                                                           & 3                                                                     & 32         & 0.05                                                     & 30    & Accuracy \\
        NUS-WIDE   & 96380                                                       & 64744                                                   & 634                                                     & 10                                                             & 5                                                            & 3                                                                     & 32         & 0.05                                                     & 20    & Accuracy \\
        Income     & 36178                                                       & 9044                                                    & 104                                                     & 10                                                             & 2                                                            & 2                                                                     & 32         & 0.05                                                     & 10    & F1-Score \\ \bottomrule
        \end{tabular}
    \label{table:1}
    \end{table*}

\section{Evaluation}

\subsection{Experimental Setup}

\subsubsection{Dataset and Training Details}
We evaluate HASSLE on four image classification datasets (\textbf{CIFAR-10, CIFAR-100\cite{krizhevsky2009learning}, ImageNette\cite{imagenette}, NUS-WIDE\cite{chua2009nus}}) and one tabular binary classification dataset (\textbf{Income\cite{income}}). \textbf{CIFAR-10} is a standard and simple dataset that contains 10 classes of RGB image samples, each of 32$\times$32 pixels. \textbf{CIFAR-100} has the same image resolution but includes 100 different classes of samples for a more challenging attack assessment. \textbf{ImageNette} is a large-scale image dataset comprising 10 easily distinguishable classes extracted from ImageNet, with the selected samples down-scaled to 128$\times$128 pixels. In our SplitVFL implementation, all samples from these three datasets are partitioned horizontally into equal size segments. Each segment is allocated to an individual party. ResNet-18 is employed as the bottom model architecture and the top model is a three-layer MLP network. \textbf{NUS-WIDE} consists of preprocessed low-level features of the collected images instead of RGB pixels. It is similar to a multi-class classification tabular dataset, which is useful for evaluating the overall attack performance. We select the 5 classes (Animals, Buildings, Grass, Person, Water) with relatively balanced number of samples for our evaluation. Since the low-level features cannot be partitioned directly, we assign the color histogram, color auto-correlogram and wavelet texture features to the active party, and edge direction histogram and block-wise color moments features to the passive party. \textbf{Income} dataset aims at predicting if a person has more than \$50000 of annual income based on tabular census data. The features are divided equally for each party. Both the bottom and top models of NUS-WIDE and Income datasets use MLP networks. 

The optimization of all the models across different datasets is performed using an SGD optimizer with a cosine annealing schedule. Unless stated otherwise, there are two parties in the VFL system. The detailed model architecture and training hyperparameters information are shown in Table \ref{table:1}.

\subsubsection{Attack Settings}

\begin{table}[b]
    \caption{LIA Attack Settings}
    \centering
    \begin{tabular}{@{}c|c|ccccc@{}}
        \toprule
        Dataset               & $r$           & \multicolumn{5}{c}{Attack Epoch ($E_a$)} \\
        \multicolumn{1}{l|}{} & \multicolumn{1}{l|}{} & Fu-LIA  & SDD  & VILLAIN  & DS  & HASSLE \\ \midrule
        CIFAR-10              & 4                     & 10      & 7    & 7        & 7   & 7      \\
        CIFAR-100             & 4                     & 20      & 13   & 13       & 13  & 13     \\
        ImageNette            & 4                     & 15      & 10   & 10       & 10  & 10     \\
        NUS-WIDE              & 4                     & 10      & 5    & 5        & 5   & 5      \\
        Income                & 8                     & 5       & 2    & 2        & 2   & 2      \\ \bottomrule
        \end{tabular}
    \label{LIA_setting}
\end{table}

The attacker is a passive party who aims at forcing the top model to output a target prediction after launching the attack at VFL test time. This target label is set to be the first class of all image classification datasets and the ``True'' class for the Income dataset. This choice is motivated by the class imbalance within the Income dataset, where samples labeled with ``True'' are much fewer than those with ``False''. Forcing a VFL model to predict a minority class is more challenging. Moreover, being predicted to have a higher income is typically more beneficial to an attacker seeking access to financial services.

Unlike previous works, we separate the LIA and hijacking attack evaluations since the LIA accuracy can have a significant impact on the hijacking attack performance. For a fair comparison among all hijacking attacks, we assume the attacker has the knowledge of $p\%$ sample IDs corresponding to the target label after the LIA. This setting eliminates the variability introduced by inconsistent LIA accuracy.

For LIA evaluation, we compare our method with Fu-LIA \cite{fu2022label}, SDD \cite{xuan2023practical}, VILLAIN \cite{bai2023villain} and Direction Score (DS). As typical hijacking attack methods only require the knowledge of a sufficient amount of target-label sample IDs, we use these LIA methods to infer a list of sample IDs with target label. To this end, the vanilla Fu-LIA is modified to select the top $\frac{n_t}{r}$ samples with the highest target-label scores as the inferred samples. We follow the original work to allow the attacker to have 50 samples per class for Income dataset and 4 known samples per class for the others. Both SDD and VILLAIN use embedding swapping technique, where the attacker swaps 5 out of 32 samples per batch. Moreover, VILLAIN doubles the learning rate of the attacker’s bottom model. Instead of manually choosing a threshold for these two LIAs, we pick the top $\frac{100}{r \times C}$ percent of samples as their inferred result. DS takes the cosine similarity score in a single epoch $E_a$ and selects the top $\frac{n_t}{r}$ samples with the highest scores. They are compared with our LIA method which takes the average scores across multiple epochs as a more robust indicator. Furthermore, our approach incorporates the SSL-based initialization for LIA evaluation. For SSL pre-training, we use MoCo-v2 \cite{chen2020improved} in the three RGB image datasets and SCARF \cite{bahri2022scarf} in the two tabular-like datasets. Since previous works have shown that more layers in the top model can reduce the gradient-based LIA accuracy, we conduct a sensitivity analysis on the LIA accuracy according to the number of top model layers. The LIA attack epoch $E_a$ and ratio $r$ for different datasets and attacks are shown in Table \ref{LIA_setting}. 

For hijacking attack evaluation, we choose LRBA\cite{gu2023lr}, Mean-Shift (MS)\cite{he2023backdoor}, BadVFL\cite{xuan2023practical} and VILLAIN\cite{bai2023villain} for comparison. Furthermore, we add a baseline attack method \textit{Replace}, which directly replaces the benign embedding with a malicious embedding generated from the known target-label sample. \textit{Grad} is our adversarial embedding attack without SSL enhancement. For LRBA, we train the attacker's surrogate model after the VFL training process with 4 known samples for each task label. We initialize the adversarial embedding to be one of the known target-label embeddings, and optimizes it using the surrogate model for 10000 steps at a step size of 1. We evaluate the optimized adversarial embedding by substituting the original benign embedding with it, without poisoning the attacker's bottom model, as this should yield a stronger attack effect. For MS, we follow the setting of the original paper and implement the mean-shift algorithm using scikit-learn package \cite{scikit-learn} in python. For BadVFL, the input trigger is a white patch for the first three image datasets. The trigger sizes are 5$\times$5, 5$\times$5 and 20$\times$20 for CIFAR-10, CIFAR-100 and ImageNette, respectively. The trigger sizes for NUS-WIDE and Income datasets are one-fifth of their respective attacker-held feature sizes, with the value set to the maximum of the corresponding dimension in the training datasets. For the embedding-level trigger in VILLAIN, we set its trigger size to 75\% of the embedding dimension with a $\gamma$ value of 3. Based on the number of revealed IDs in our LIA, the poisoning ratios are set to 2.5\%, 0.25\%, 2.5\%, 5\% and 6.25\% for CIFAR-10, CIFAR-100, ImageNette, NUS-WIDE and Income, respectively across all the attacks. 

\subsubsection{Evaluation Metrics}
LIAs generate a list of sample IDs that are inferred to have a target label. Precision, which is the ratio of the true positive counts to all positive counts, is used to quantify their performance. The precision of a good LIA should be close to 1. On the other hand, the hijacking attack performance is measured by the ASR, which is the percentage of test samples predicted as the target label under the attack. Finally, the stealthiness of the hijacking attack can be assessed by the difference in the main task accuracy (MTA) between an attacked model and an unscathed model.

\begin{table}[t]
    \centering
    \caption{Comparison of the precision (in \%) of different LIA methods with different number of top model layers.}
    \label{LIA}
    \begin{tabular}{@{}cccccc@{}}
        \toprule
        \multicolumn{6}{c}{CIFAR-10}                                          \\
        \multicolumn{1}{c|}{Layers} & Fu-LIA & SDD  & VILLAIN & DS   & HASSLE (Ours) \\ \midrule
        \multicolumn{1}{c|}{1}      & 99.5   & 100  & 10.1    & 100  & 100    \\
        \multicolumn{1}{c|}{2}      & 99.7   & 93   & 16.1    & 100  & 99.4   \\
        \multicolumn{1}{c|}{3}      & 99.1   & 75.1 & 15.4    & 100  & 100    \\
        \multicolumn{1}{c|}{4}      & 98.2   & 100  & 9.3     & 97.3 & 99.7   \\
        \multicolumn{1}{c|}{5}      & 98.6   & 86.2 & 13.2    & 97.5 & 99.2   \\ \midrule
        \multicolumn{6}{c}{CIFAR-100}                                         \\
        \multicolumn{1}{c|}{Layers} & Fu-LIA & SDD  & VILLAIN & DS   & HASSLE (Ours) \\ \midrule
        \multicolumn{1}{c|}{1}      & 98.4   & 100  & 6       & 100  & 100    \\
        \multicolumn{1}{c|}{2}      & 87.2   & 44.4 & 3       & 93.6 & 100    \\
        \multicolumn{1}{c|}{3}      & 100    & 1.8  & 1.5     & 88   & 100    \\
        \multicolumn{1}{c|}{4}      & 91.2   & 12.9 & 2       & 48.8 & 84     \\
        \multicolumn{1}{c|}{5}      & 1      & 55.6 & 9       & 59.2 & 41.6   \\ \midrule
        \multicolumn{6}{c}{ImageNette}                                        \\
        \multicolumn{1}{c|}{Layers} & Fu-LIA & SDD  & VILLAIN & DS   & HASSLE (Ours) \\ \midrule
        \multicolumn{1}{c|}{1}      & 100    & 100  & 4       & 100  & 100    \\
        \multicolumn{1}{c|}{2}      & 100    & 100  & 7.3     & 99.2 & 100    \\
        \multicolumn{1}{c|}{3}      & 100    & 94.6 & 10.8    & 100  & 100    \\
        \multicolumn{1}{c|}{4}      & 100    & 86.5 & 13.6    & 98.7 & 100   \\
        \multicolumn{1}{c|}{5}      & 51.27  & 86.5 & 18.2    & 54.7 & 100    \\ \midrule
        \multicolumn{6}{c}{NUS-WIDE}                                          \\
        \multicolumn{1}{c|}{Layers} & Fu-LIA & SDD  & VILLAIN & DS   & HASSLE (Ours) \\ \midrule
        \multicolumn{1}{c|}{1}      & 77.4   & 100  & 18.1    & 100  & 100    \\
        \multicolumn{1}{c|}{2}      & 80.8   & 100  & 19.1    & 100  & 100    \\
        \multicolumn{1}{c|}{3}      & 78.1   & 99.5 & 14.6    & 86.9 & 100    \\
        \multicolumn{1}{c|}{4}      & 73.1   & 98.5 & 16.7    & 98.8 & 100    \\
        \multicolumn{1}{c|}{5}      & 100    & 77.7 & 15.3    & 90.9 & 100    \\ \midrule
        \multicolumn{6}{c}{Income}                                            \\
        \multicolumn{1}{c|}{Layers} & Fu-LIA & SDD  & VILLAIN & DS   & HASSLE (Ours) \\ \midrule
        \multicolumn{1}{c|}{1}      & 73.2   & 100  & 0.9     & 100  & 100    \\
        \multicolumn{1}{c|}{2}      & 77.4   & 100  & 0.4     & 100  & 100    \\
        \multicolumn{1}{c|}{3}      & 90.8   & 100  & 2.6     & 100  & 100    \\
        \multicolumn{1}{c|}{4}      & 36.5   & 100  & 3.6     & 100  & 100    \\ \bottomrule
        \end{tabular}
\end{table}

\subsection{Results}
\subsubsection{LIA Performance} \label{label_inference_analysis}
\begin{figure}[t]
    \centering
    \includegraphics[width=0.7\linewidth]{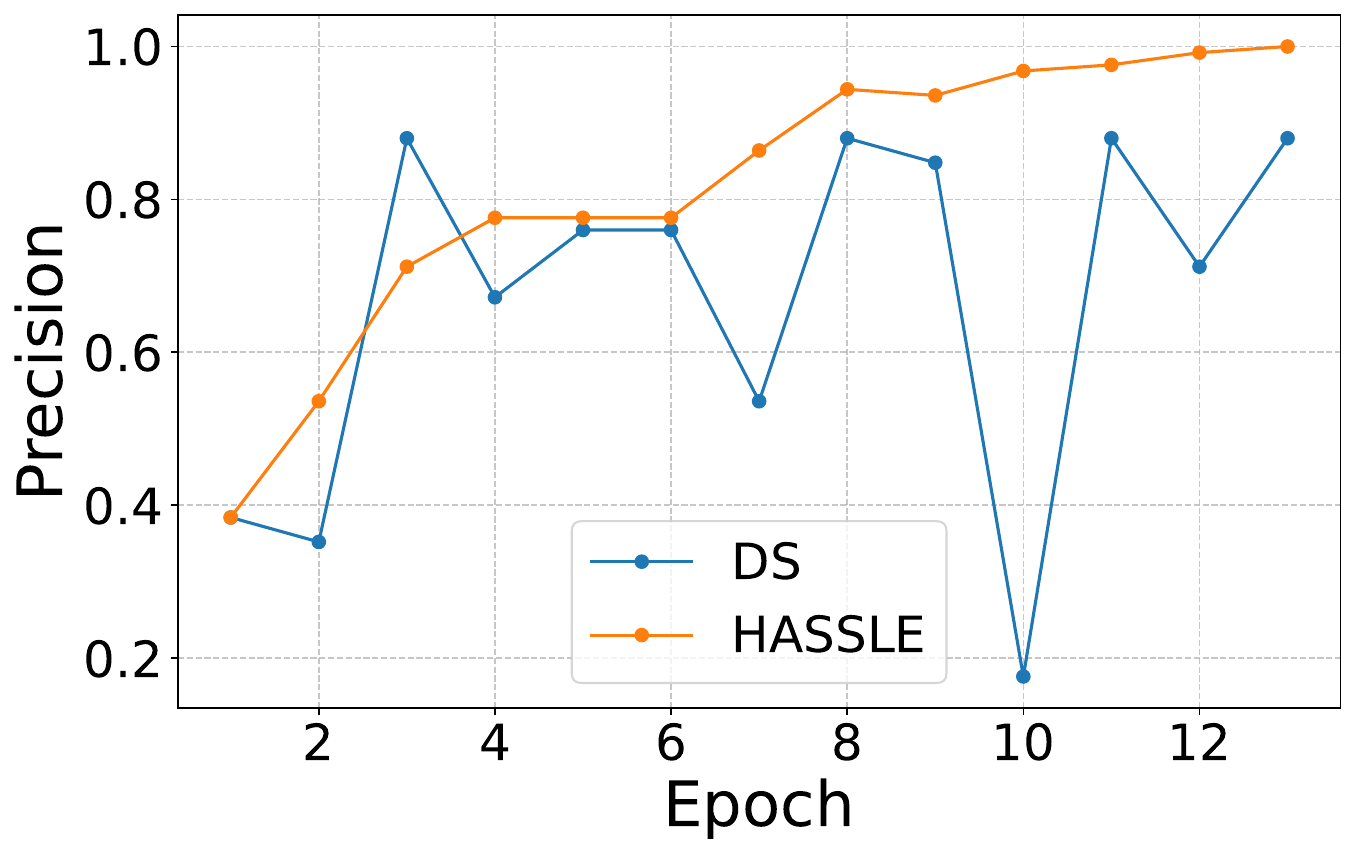}
    \caption{Comparison of the LIA precision in each epoch before $E_a$ of DS and HASSLE for CIFAR-100 dataset.}
    \label{ds_hassle}
\end{figure}

As shown in Table \ref{LIA}, HASSLE LIA achieves superior inference precision compared to existing methods. Conversely, VILLAIN performs poorly on all datasets with an attack precision close to or even worse than the random guessing. This is because the embedding swapping module in VILLAIN does not function as effectively as claimed. The purpose of embedding swapping is to confuse the top model prediction when the swapped sample corresponds to a non-target label, while keeping the prediction intact when swapping a target-label sample. It assumes that the returned gradient norm will be higher for non-target labels and lower for target labels. However, due to the non-linearity of the top model's MLP, embedding swapping does not reliably produce the desired changes in the top model's predictions. On one hand, the top model may still produce a similar prediction score to the unswapped one for the sample with non-target label. On the other hand, swapping embeddings for target-label samples can lead to less confident predictions for the target label because the top model has not been trained on such swapped combinations. Although embedding swapping is also employed in SDD, its attack precision is much higher than VILLAIN except for the more challenging CIFAR-100. This discrepancy is likely due to the weak correlation between the gradient direction and the label as explained in Section \ref{3-c}, despite the swap incurs perturbation in model prediction score. DS achieves much higher precision without being affected by embedding swapping, but there is an epoch of instability arising from the stochastic update of model parameters. Our HASSLE LIA mitigates the fluctuation of DS precision as shown in Fig. \ref{ds_hassle} through averaging the similarity scores across multiple epochs. Fu-LIA requires knowledge of substantially more training samples than the others and relies on a powerful bottom model. It attains perfect precision in the three datasets using more sophisticated bottom model architecture but performs poorly in the two that use simple MLP model. This performance drop also results from the class imbalance in NUS-WIDE and Income datasets, which distorts the pseudo labels generated during its semi-supervised learning process.

Our LIA method consistently achieves more than 90\% of precision regardless of the number of top model layers, except on CIFAR-100 for more than three layers. In contrast, other attacks stumble when the number of layers increases. All LIA methods suffer precision degradation on CIFAR-100 dataset since its number of classes is much larger than other datasets, making the target label identification more difficult. Moreover, the number of training samples for each class in CIFAR-100 is far from enough to train a capable classification model, which denies the orthogonality of column vectors in the weight matrix, a property necessary for preserving the correlation between the gradient direction and the label. For Income, all three gradient-direction-based LIA methods achieve perfect precision across different number of top model layers, which matches our expectation that gradient direction can easily leak private labels in VFL binary classification.

\begin{table*}[t]
    \centering
    \caption{Comparison of Attack Success Rate (ASR) and Main Task Accuracy (MTA) (both in \%) of Hijacking Attack Methods. \textbf{Bold} indicates the best-performing method.}
    \label{hijack}
    \begin{subtable}{0.95\linewidth}
    \centering
    \begin{tabular}{@{}c|cc|cc|c|cc|cc|cc|cc|cc@{}}
    \toprule
                                         & \multicolumn{2}{c|}{\textbf{Replace}}         & \multicolumn{2}{c|}{\textbf{LRBA}}            & \textbf{} & \multicolumn{2}{c|}{\textbf{MS}} & \multicolumn{2}{c|}{\textbf{BadVFL}} & \multicolumn{2}{c|}{\textbf{VILLAIN}} & \multicolumn{2}{c|}{\textbf{Grad}} & \multicolumn{2}{c}{\textbf{HASSLE}} \\
                                         & ASR                   & MTA                   & ASR                   & MTA                   & $E_a$     & ASR             & MTA            & ASR               & MTA              & ASR               & MTA               & ASR              & MTA             & ASR                 & MTA           \\ \midrule
    \multirow{3}{*}{\textbf{CIFAR-10}}   & \multirow{3}{*}{44.9} & \multirow{3}{*}{81.1} & \multirow{3}{*}{78.4} & \multirow{3}{*}{81.1} & 5         & 68.7            & 81.1           & 86.7              & 81.3             & 71.5              & 81.7              & 77.9             & 81.3            & \textbf{99.6}          & 83.43         \\
                                         &                       &                       &                       &                       & 7         & 62.1            & 81             & 78.4              & 81.9             & 67.1              & 81.4              & 81.8             & 81.3            & \textbf{99.8}          & 83.2          \\
                                         &                       &                       &                       &                       & 9         & 51.3            & 80.9           & 77.8              & 81.3             & 40.1              & 82                & 67.3             & 81.4            & \textbf{99.2}          & 83.3          \\ \midrule
    \multirow{3}{*}{\textbf{CIFAR-100}}  & \multirow{3}{*}{43.2} & \multirow{3}{*}{47.4} & \multirow{3}{*}{0.6}  & \multirow{3}{*}{47.4} & 10        & 9.7             & 46.8           & 68.2              & 47.7             & 7.7               & 47.8              & 54.7             & 47.5            & \textbf{86.3}          & 52.1          \\
                                         &                       &                       &                       &                       & 13        & 12.1            & 46.8           & 54                & 47.3             & 4.6               & 47.5              & 48.6             & 47.3            & \textbf{85}            & 51.5          \\
                                         &                       &                       &                       &                       & 16        & 5.7             & 47             & 61.5              & 47.1             & 2.5               & 47.7              & 46.9             & 47.6            & \textbf{83.5}          & 51.3          \\ \midrule
    \multirow{3}{*}{\textbf{ImageNette}} & \multirow{3}{*}{74.4} & \multirow{3}{*}{75.5} & \multirow{3}{*}{76.8} & \multirow{3}{*}{75.5} & 7         & 1.9             & 75.8           & 65.2              & 75.8             & 48.7              & 74.1              & 84.1             & 75.7            & \textbf{99.8}          & 78.6          \\
                                         &                       &                       &                       &                       & 10        & 5.2             & 74.9           & 69.5              & 74.6             & 48.7              & 75                & 85.6             & 75.8            & \textbf{99.3}          & 78.9          \\
                                         &                       &                       &                       &                       & 13        & 1.6             & 74.8           & 63.8              & 75.4             & 26.9              & 75                & 75.5             & 75.8            & \textbf{100}           & 79.2          \\ \midrule
    \multirow{3}{*}{\textbf{NUS-WIDE}}   & \multirow{3}{*}{14.6} & \multirow{3}{*}{65.7} & \multirow{3}{*}{59.2} & \multirow{3}{*}{65.7} & 3         & 99.9            & 65.4           & \textbf{100}               & 65.7             & 94.6              & 65.6              & 96.9             & 65.6            & \textbf{100}           & 66.4          \\
                                         &                       &                       &                       &                       & 5         & 99.8            & 65.4           & \textbf{100}               & 65.7             & 94.3              & 65.5              & 96.7             & 65.5            & 99.8          & 66.4          \\
                                         &                       &                       &                       &                       & 7         & 99.7            & 65.2           & \textbf{100}               & 65.7             & 94.2              & 65.7              & 96.1             & 65.6            & \textbf{100}           & 66.3          \\ \midrule
    \multirow{3}{*}{\textbf{Income}}     & \multirow{3}{*}{6.1}  & \multirow{3}{*}{69.6} & \multirow{3}{*}{5.6}  & \multirow{3}{*}{69.6} & 1         & 99.9            & 65.9           & \textbf{100}               & 66.5             & 96.1              & 66.8              & 97.9             & 67              & \textbf{100}           & 65.7          \\
                                         &                       &                       &                       &                       & 2         & 99.9            & 65.6           & \textbf{100}               & 66               & 96.1              & 66.6              & 96.6             & 66.4            & \textbf{100}           & 65.4          \\
                                         &                       &                       &                       &                       & 3         & \textbf{100}             & 66.2           & \textbf{100}               & 67.1             & 96.2              & 67.6              & 94.3             & 67.3            & \textbf{100}           & 66            \\ \bottomrule
    \end{tabular}
    \vspace{3pt}
    \caption{Number of Parties = 2}
    \end{subtable}

    \begin{subtable}{0.95\linewidth}
    \centering
    \begin{tabular}{@{}c|cc|cc|c|ll|ll|ll|ll|ll@{}}
        \toprule
                                             & \multicolumn{2}{c|}{\textbf{Replace}}         & \multicolumn{2}{c|}{\textbf{LRBA}}            & \textbf{} & \multicolumn{2}{c|}{\textbf{MS}}                   & \multicolumn{2}{c|}{\textbf{BadVFL}}               & \multicolumn{2}{c|}{\textbf{VILLAIN}}              & \multicolumn{2}{c|}{\textbf{Grad}}                 & \multicolumn{2}{c}{\textbf{HASSLE}}               \\
                                             & ASR                   & MTA                   & ASR                   & MTA                   & $E_a$     & \multicolumn{1}{c}{ASR} & \multicolumn{1}{c|}{MTA} & \multicolumn{1}{c}{ASR} & \multicolumn{1}{c|}{MTA} & \multicolumn{1}{c}{ASR} & \multicolumn{1}{c|}{MTA} & \multicolumn{1}{c}{ASR} & \multicolumn{1}{c|}{MTA} & \multicolumn{1}{c}{ASR} & \multicolumn{1}{c}{MTA} \\ \midrule
        \multirow{3}{*}{\textbf{CIFAR-10}}   & \multirow{3}{*}{14.1} & \multirow{3}{*}{79.1} & \multirow{3}{*}{30.8} & \multirow{3}{*}{79.1} & 5         & 56.2                    & 78.3                     & 62.5                    & 78.4                     & 40.9                    & 78.5                     & 67.9                    & 78.6                     & \textbf{90.8}                    & 79.4                    \\
                                             &                       &                       &                       &                       & 7         & 45.3                    & 78.2                     & 44.2                    & 78.1                     & 24.5                    & 78.7                     & 58.8                    & 78.2                     & \textbf{79.4}                    & 79.4                    \\
                                             &                       &                       &                       &                       & 9         & 47.0                    & 78.5                     & 43.7                    & 78.5                     & 13.3                    & 78.4                     & 56.3                    & 78.2                     & \textbf{77.6}                    & 79.4                    \\ \midrule
        \multirow{3}{*}{\textbf{CIFAR-100}}  & \multirow{3}{*}{38.8} & \multirow{3}{*}{45.8} & \multirow{3}{*}{3.1}  & \multirow{3}{*}{45.8} & 10        & 6.4                     & 46.2                     & 67.2                    & 46.2                     & 9.2                     & 46.5                     & 63.4                    & 46.2                     & \textbf{81.8}                    & 49.9                    \\
                                             &                       &                       &                       &                       & 13        & 12.6                    & 46.5                     & 64.7                    & 46.9                     & 3.7                     & 47.1                     & 44.3                    & 46.2                     & \textbf{72.4}                    & 49.5                    \\
                                             &                       &                       &                       &                       & 16        & 12.7                    & 46.5                     & 59.3                    & 46.7                     & 4.7                     & 46.8                     & 43.0                    & 45.8                     & \textbf{65.6}                    & 49.5                    \\ \midrule
        \multirow{3}{*}{\textbf{ImageNette}} & \multirow{3}{*}{51.7} & \multirow{3}{*}{76.7} & \multirow{3}{*}{96.4} & \multirow{3}{*}{76.7} & 7         & 94.8                    & 75.5                     & 84.2                    & 75.1                     & 59.6                    & 74.9                     & 93.8                    & 75.2                     & \textbf{100}                   & 78.8                    \\
                                             &                       &                       &                       &                       & 10        & 89.2                    & 75.2                     & 79.4                    & 75.2                     & 41.9                    & 75.4                     & 98.8                    & 74.6                     & \textbf{100}                   & 78.6                    \\
                                             &                       &                       &                       &                       & 13        & 95.8                    & 75.2                     & 81.2                    & 75.4                     & 27.0                    & 75.4                     & 93.2                    & 76.2                     & \textbf{100}                   & 78.6                    \\ \midrule
        \multirow{3}{*}{\textbf{NUS-WIDE}}   & \multirow{3}{*}{5.3}  & \multirow{3}{*}{66.3} & \multirow{3}{*}{\textbf{100}}  & \multirow{3}{*}{66.3} & 3         & 98.3                    & 65.9                     & \textbf{100}                   & 65.9                     & 99.9                    & 65.8                     & 99.5                    & 66.0                     & 99.7                    & 65.6                    \\
                                             &                       &                       &                       &                       & 5         & 96.5                    & 65.8                     & \textbf{100}                   & 65.8                     & 99.9                    & 65.8                     & 99.0                    & 66.0                     & 99.8                    & 65.7                    \\
                                             &                       &                       &                       &                       & 7         & 94.9                    & 65.9                     & \textbf{100}                   & 65.8                     & 99.8                    & 65.8                     & 98.6                    & 65.8                     & 99.6                    & 65.6                    \\ \midrule
        \multirow{3}{*}{\textbf{Income}}     & \multirow{3}{*}{22.8} & \multirow{3}{*}{69.5} & \multirow{3}{*}{25.7} & \multirow{3}{*}{69.5} & 1         & \textbf{99.9}                    & 65.2                     & 99.2                    & 65.6                     & 94.4                    & 65.6                     & 99.5                    & 65.8                     & \textbf{99.9}                    & 65.8                    \\
                                             &                       &                       &                       &                       & 2         & \textbf{99.9}                    & 65.5                     & 98.9                    & 65.8                     & 94.1                    & 65.8                     & 99.5                    & 65.6                     & \textbf{99.9}                    & 65.9                    \\
                                             &                       &                       &                       &                       & 3         & \textbf{99.8}                    & 64.9                     & 97.0                    & 65.6                     & 93.6                    & 65.7                     & 98.6                    & 65.2                     & \textbf{99.8}                    & 65.9                    \\ \bottomrule
        \end{tabular}
        \vspace{3pt}
        \caption{Number of Parties = 3}
    \end{subtable}

    \begin{subtable}{0.95\linewidth}
    \centering        
    \begin{tabular}{@{}c|cc|cc|c|ll|ll|ll|ll|ll@{}}
        \toprule
                                             & \multicolumn{2}{c|}{\textbf{Replace}}         & \multicolumn{2}{c|}{\textbf{LRBA}}            & \textbf{} & \multicolumn{2}{c|}{\textbf{MS}}                   & \multicolumn{2}{c|}{\textbf{BadVFL}}               & \multicolumn{2}{c|}{\textbf{VILLAIN}}              & \multicolumn{2}{c|}{\textbf{Grad}}                 & \multicolumn{2}{c}{\textbf{HASSLE}}               \\
                                             & ASR                   & MTA                   & ASR                   & MTA                   & $E_a$     & \multicolumn{1}{c}{ASR} & \multicolumn{1}{c|}{MTA} & \multicolumn{1}{c}{ASR} & \multicolumn{1}{c|}{MTA} & \multicolumn{1}{c}{ASR} & \multicolumn{1}{c|}{MTA} & \multicolumn{1}{c}{ASR} & \multicolumn{1}{c|}{MTA} & \multicolumn{1}{c}{ASR} & \multicolumn{1}{c}{MTA} \\ \midrule
        \multirow{3}{*}{\textbf{CIFAR-10}}   & \multirow{3}{*}{13.9} & \multirow{3}{*}{76.3} & \multirow{3}{*}{11.3} & \multirow{3}{*}{76.3} & 5         & 39.6                    & 76.2                     & 48.3                    & 75.9                     & 42.0                    & 76.3                     & 63.9                    & 75.7                     & \textbf{67.0}                    & 75.1                    \\
                                             &                       &                       &                       &                       & 7         & 26.3                    & 75.6                     & 34.3                    & 75.9                     & 22.7                    & 76.5                     & 52.1                    & 75.5                     & \textbf{52.9}                    & 76.0                    \\
                                             &                       &                       &                       &                       & 9         & 16.8                    & 75.9                     & 18.8                    & 75.8                     & 11.1                    & 76.5                     & 22.0                    & 75.4                     & \textbf{41.0}                    & 75.6                    \\ \midrule
        \multirow{3}{*}{\textbf{CIFAR-100}}  & \multirow{3}{*}{4}    & \multirow{3}{*}{43.8} & \multirow{3}{*}{7.5}  & \multirow{3}{*}{43.8} & 10        & 18.0                    & 43.7                     & 43.4                    & 44.2                     & 4.5                     & 43.9                     & 45.7                    & 43.7                     & \textbf{47.3}                    & 44.8                    \\
                                             &                       &                       &                       &                       & 13        & 1.9                     & 44.0                     & 43.5                    & 43.8                     & 2.5                     & 43.8                     & 28.4                    & 43.9                     & \textbf{44.4}                    & 45.6                    \\
                                             &                       &                       &                       &                       & 16        & 2.7                     & 43.9                     & \textbf{27.9}                   & 43.3                     & 1.6                     & 43.6                     & 16.6                    & 43.8                     & 22.8                    & 45.0                    \\ \midrule
        \multirow{3}{*}{\textbf{ImageNette}} & \multirow{3}{*}{5}    & \multirow{3}{*}{71.7} & \multirow{3}{*}{71.9} & \multirow{3}{*}{71.7} & 7         & 1.2                     & 70.7                     & 36.2                    & 70.1                     & 53.1                    & 70.8                     & 63.1                    & 70.4                     & \textbf{93.9}                    & 71.5                    \\
                                             &                       &                       &                       &                       & 10        & 0.5                     & 69.2                     & 26.9                    & 70.6                     & 42.9                    & 70.8                     & 57.7                    & 70.1                     & \textbf{91.6}                    & 71.7                    \\
                                             &                       &                       &                       &                       & 13        & 0.3                     & 69.8                     & 15.7                    & 70.6                     & 25.0                    & 71.3                     & 28.5                    & 70.3                     & \textbf{88.8}                    & 71.0                    \\ \midrule
        \multirow{3}{*}{\textbf{NUS-WIDE}}   & \multirow{3}{*}{1.4}  & \multirow{3}{*}{66.2} & \multirow{3}{*}{22.3} & \multirow{3}{*}{66.2} & 3         & 99.5                    & 65.7                     & \textbf{100}                   & 65.6                     & 99.7                    & 65.6                     & 98.5                    & 65.7                     & 97.3                    & 66.1                    \\
                                             &                       &                       &                       &                       & 5         & 99.2                    & 65.7                     & \textbf{99.9}                    & 65.8                     & 99.7                    & 65.8                     & 97.9                    & 65.7                     & 96.5                    & 66.1                    \\
                                             &                       &                       &                       &                       & 7         & 98.1                    & 65.7                     & \textbf{99.8}                    & 65.8                     & 99.4                    & 66.0                     & 96.3                    & 65.6                     & 95.7                    & 66.2                    \\ \midrule
        \multirow{3}{*}{\textbf{Income}}     & \multirow{3}{*}{26.7} & \multirow{3}{*}{69.1} & \multirow{3}{*}{31.4} & \multirow{3}{*}{69.1} & 1         & \textbf{100}                   & 66.2                     & 99.3                    & 66.5                     & 95.3                    & 66.6                     & 99.7                    & 66.4                     & \textbf{100}                   & 65.3                    \\
                                             &                       &                       &                       &                       & 2         & \textbf{100}                   & 66.5                     & 98.9                    & 66.5                     & 94.6                    & 66.6                     & 99.6                    & 66.3                     & \textbf{100}                   & 65.2                    \\
                                             &                       &                       &                       &                       & 3         & 99.9                    & 66.0                     & 98.7                    & 66.4                     & 93.3                    & 66.2                     & 99.1                    & 66.0                     & \textbf{100}                   & 65.1                    \\ \bottomrule
        \end{tabular}
        \vspace{3pt}
        \caption{Number of Parties = 4}
    \end{subtable}
    \vspace{-20pt}
    \end{table*}

\subsubsection{Hijacking Attack Performance}
Table \ref{hijack} presents the effectiveness of hijacking attacks and main task performance upon completion of the VFL training for different attack launching epoch ($E_a$) and number of parties. We start by analyzing the two-party scenario. Notably, the MTA under the \textit{Replace} attack represents the original model utility without any training-time disruption. We can directly observe that this naive attack can already reach above 40\% ASR for the image datasets, though it remains relatively weak in the two tabular datasets possibly due to the weaker feature saliency held by the attacker's embeddings. LRBA and MS have inconsistent performance across different datasets. This is because LRBA relies on a surrogate model highly similar to the actual top model which is challenging to achieve in difficult datasets, and MS leverages mean-shift algorithm that has a high performance variance. While BadVFL and VILLAIN both show greater attack stability, BadVFL performs constantly better than the VILLAIN attack. We attribute this difference to the nature of their triggers: the input-level trigger in BadVFL naturally integrates with target-label embeddings, which offers better learnability for the top model, while the additive embedding-level trigger in VILLAIN is more challenging for the top model to learn effectively. The attack performance of our proposed \textit{Grad} attack is on-par with BadVFL for the CIFAR-10 and ImageNette datasets, but slightly worse than BadVFL for the CIFAR-100, NUS-WIDE and Income datasets. After applying the SSL-based initialization, our full-fledged HASSLE attack boosts its ASR to almost 100\% for all datasets except CIFAR-100, where it can still attain above 80\% ASR. While there is no evident MTA drop for any attacks, we notice a minor improvement on MTA in HASSLE attack. This suggests that the VFL model performance can be benefited when its bottom models are first trained to learn meaningful representations. For different $E_a$, all the attacks generally follow a trend that an earlier launching epoch results in a higher ASR. This can be intuitively understood that the top model has more iterations to learn the adversarial feature and associate it with the target label. However, the attacker needs a $E_a$ large enough to achieve high LIA precision as mentioned previously.

For the scenarios where the number of parties is more than two, we divide the original samples in the dataset along the input feature dimension into equal partitions by the total number of parties. Each passive party sends an embedding with the same dimension for each training sample ID to the active party for concatenation. The attacker therefore has less input features and top model prediction influence. Expectedly, the result shows that all evaluated attacks suffer a certain degree of ASR drop. HASSLE suffers a substantial ASR degradation when the number of parties reaches 4. One reason is the diluted poisoning effect with the attacker embedding portion reduced in the concatenation. Another possible reason is the reduced amount of input features cannot effectively utilize the SSL to produce a good bottom model initialization. The cropped image samples in CIFAR-10 and CIFAR-100 only have a pixel size of $32\times8$ in a four-party setting,  whereas the ImageNette samples have a pixel size of $32\times128$. This explains why ImageNette has a milder ASR drop. Overall, HASSLE still has decent performance for larger number of participating parties as long as the attacker-held features supports a solid SSL training, and it remains the most effective attack.

\begin{figure}[t]
    \centering
    \includegraphics[width=\linewidth]{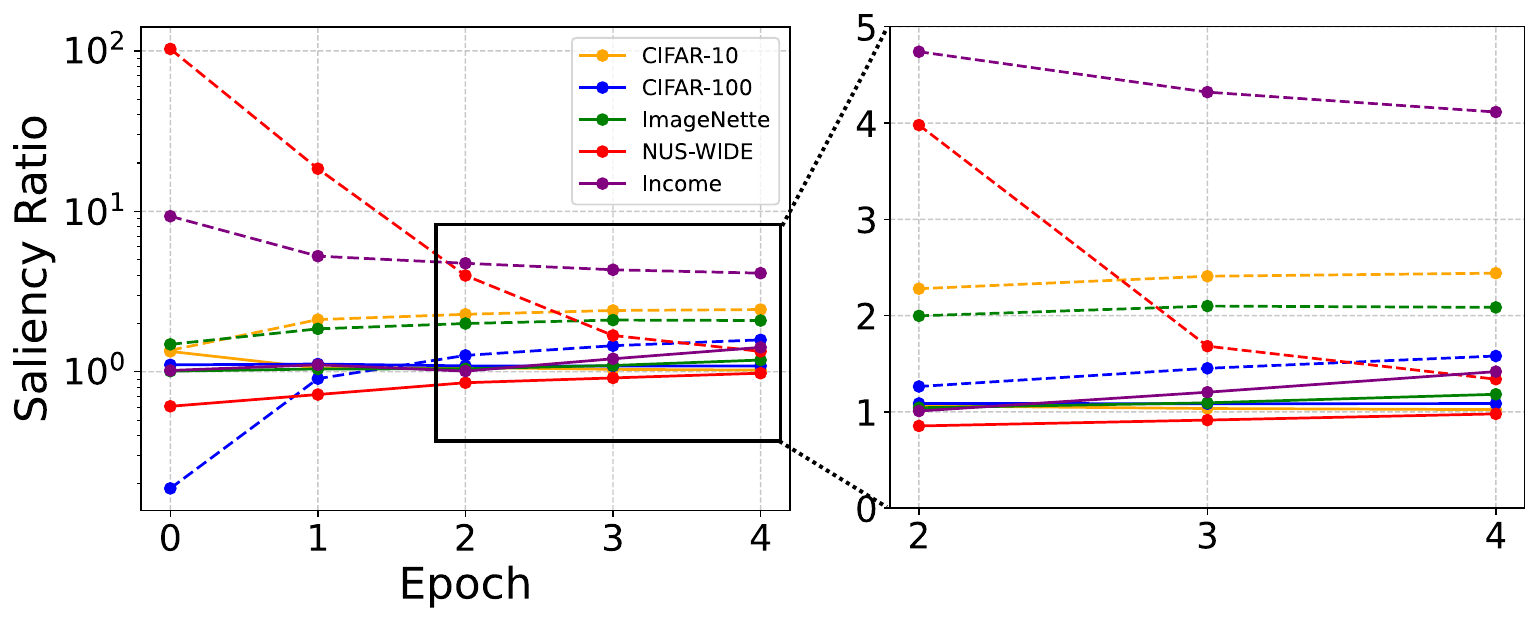}
    \caption{The feature saliency ratio between the passive and active parties under SSL initialization (dashed line) and random initialization (solid line).}
    \label{sal_ratio}
\end{figure}

We further study the effect of SSL initialization in HASSLE by comparing the feature saliency ratio between the adversarial and active parties under different bottom model initializations, without injecting adversarial embeddings. Feature saliency for each data point is computed similarly to Grad-CAM \cite{Selvaraju_2017_ICCV}, by multiplying the neuron activations to its gradient with respect to the ground truth label logit. The exact formula is shown below, where $h_k$ denotes the submitted embedding from the $k$-th party and $\odot$ is the element-wise multiplication.

\vspace{-6pt}

\begin{equation}
    \emph{Feature}\_\emph{Saliency} = \| \frac{\partial l_{gt}}{\partial h_k} \odot h_k \| \tag{11}
\end{equation}

\vspace{-5pt}

Fig. \ref{sal_ratio} shows that with SSL initialization, the adversarial party's embedding consistently exhibits higher saliency than the active party's after training stabilizes across all five datasets. Notably, while the saliency ratio decreases in the two tabular datasets, it increases in the image datasets as training proceeds. We attribute this discrepancy to differences in raw input feature information density: image objects are commonly shared between parties, whereas tabular features rarely provide equal amount of useful information. This effect is further accentuated by the faster convergence of MLP networks on tabular data. Another interesting finding is the SSL-initialized adversarial party starts with a near-zero saliency ratio in the initial epoch on CIFAR-100. This likely reflects the challenge of transferring a pretrained encoder to the CIFAR-100 classification task. It eventually surpasses the random initialization as training stabilizes with a better learned representation.

\vspace{-10pt}

\subsection{Impact of Attack Hyperparameters and VFL Settings}
In the following analysis, we launch the complete HASSLE framework by combining both its LIA and hijacking attack modules. Therefore, a low LIA precision can affect the hijacking attack performance in these evaluations.

\begin{figure}[h]
    \centering
    \includegraphics[width=0.95\linewidth]{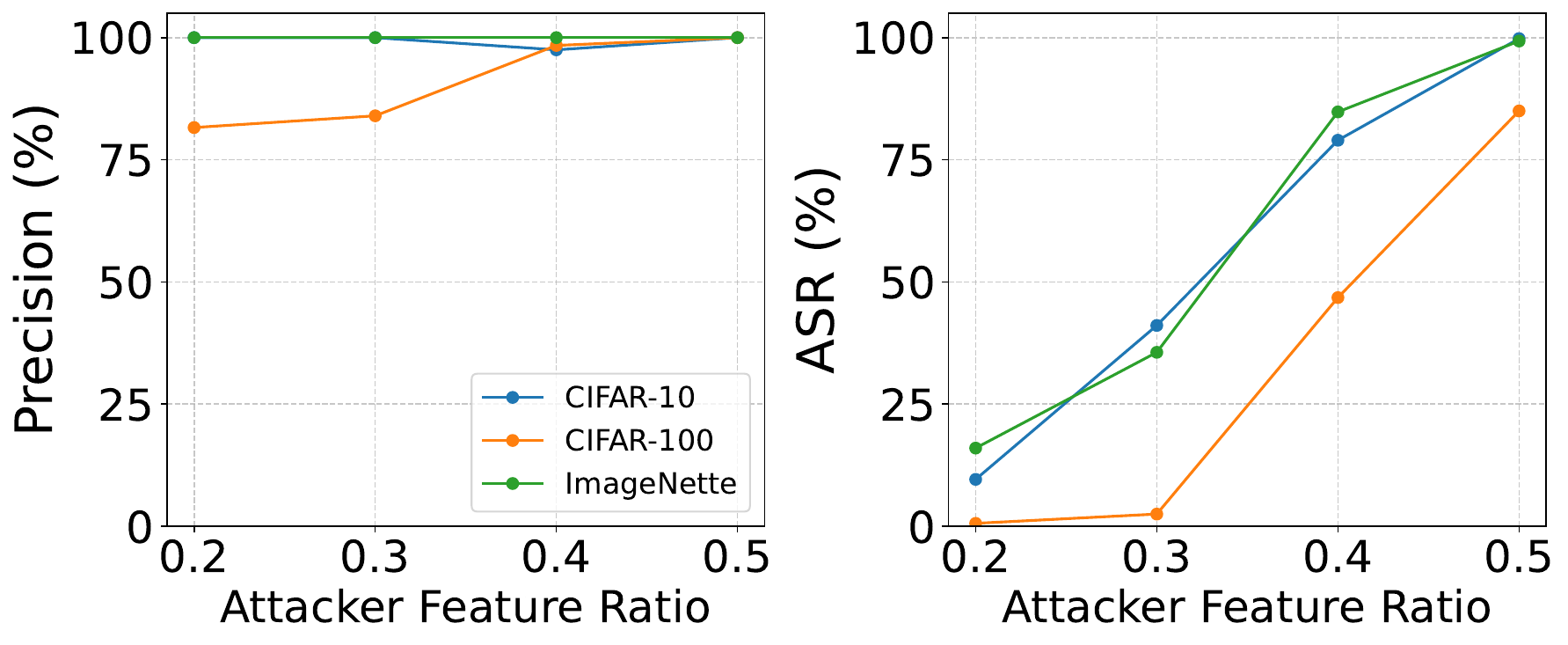}
    \caption{LIA precision and ASR of HASSLE under different attacker-held feature ratios}
    \label{feature_ratio}
\end{figure}

\subsubsection{Attacker-held Feature Ratio}
We study the impact of an unequal feature ratio distribution between the two parties on HASSLE attack performance. Since partitioning the input features of NUS-WIDE and Income datasets according to a fixed value does not provide any practical meaning, we conduct this study only on the three RGB image datasets by slicing the image samples vertically such that the attacker owns the part on the right with $\emph{Feature Ratio} \%$ of the original width. Moreover, we adjust the dimension of the submitted embeddings from the two parties such that the attacker embedding now accounts for $\emph{Feature Ratio} \%$ of the concatenated embedding at the input of the top model. Similar to the effect brought by the increasing number of parties, the attacker loses influence in the top model prediction and has suboptimal SSL-based initialization if the feature ratio becomes too low. This is illustrated in Fig.  \ref{feature_ratio}, which shows that HASSLE does not perform well in all datasets if the attacker-held feature ratio is below 0.3. More importantly, the ASR on ImageNette drops below 50\% unlike in the case of more parties where it still maintains above 80\% ASR. This observation implies that the top model relies predominantly on the greater proportion of concatenated embedding dimension held by the active party, and it is difficult for a hijacking attack to neutralize this effect.

\begin{figure}[t]
    \centering
    \includegraphics[width=0.95\linewidth]{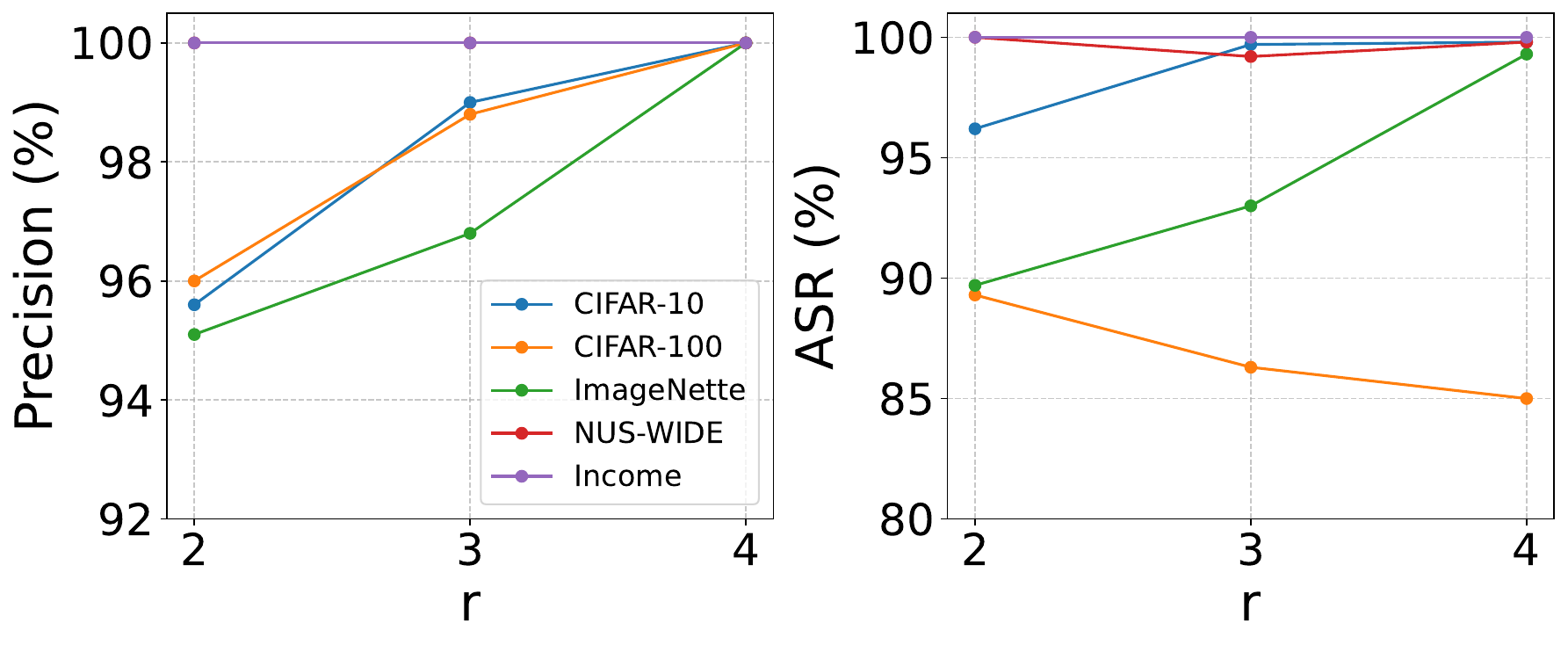}
    \caption{LIA precision and ASR of HASSLE under different $r$}
    \label{tg_ratio}
\end{figure}

\subsubsection{Varying $r$}
A larger filtering ratio $r$ improves the precision of our LIA as it only selects samples with the highest similarity scores. Fig.  \ref{tg_ratio} shows that LIA precision increases as $r$ becomes larger. A better LIA precision leads to a stronger hijacking attack as shown by the CIFAR-10 and ImageNette curves. However, a larger $r$ does not always enhance the ASR since it reduces the available training samples that can be used for poisoning, which can be observed in the CIFAR-100 ASR curve. This stark contrast is mainly due to the difference in poisoning ratio among evaluated datasets. When $r=4$, the poisoning ratio for CIFAR-100 is only 0.25\%, compared to 2.5\% for CIFAR-10 and ImageNette. Typically, a backdoor poisoning attack requires at least 1\% of training samples to be poisoned to achieve a dominant ASR. Therefore, it is beneficial to choose a smaller $r$ for a dataset with a tiny portion of target-label samples. Contrarily, for datasets with few classes like CIFAR-10 and ImageNette, the decrease in LIA precision indicates a considerable amount of poisoned samples actually belong to non-target classes. Not only does this distract the top model poisoning by associating the adversarial embedding with a non-target label, but the embedding is also optimized towards the wrong decision region. Thus, we recommend choosing a $r$ that corresponds to a poisoning ratio of 1\%.

\begin{figure}[h]
    \centering
    \includegraphics[width=0.95\linewidth]{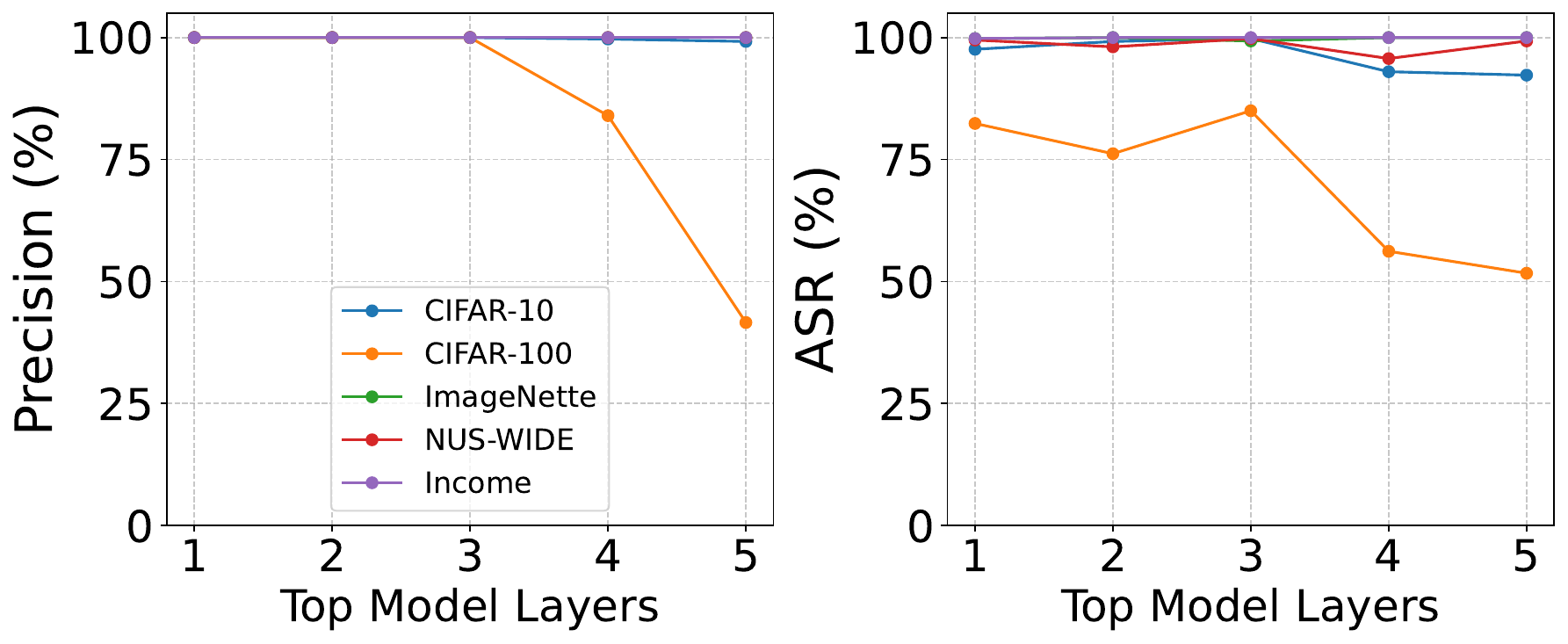}
    \caption{LIA precision and ASR of HASSLE under different number of top model layers}
    \label{num_layer}
\end{figure}

\subsubsection{Number of Top Model Layers}
As shown in Fig.  \ref{num_layer}, the number of layers in the top model has negligible influence on our hijacking attack. The decline in ASR on the CIFAR-100 dataset is mainly caused by the precision drop in the LIA module. This is due to the complexity and sample deficiency of the CIFAR-100 dataset as explained in Section \ref{label_inference_analysis}. 

\begin{figure}[h]
    \centering
    \includegraphics[width=0.95\linewidth]{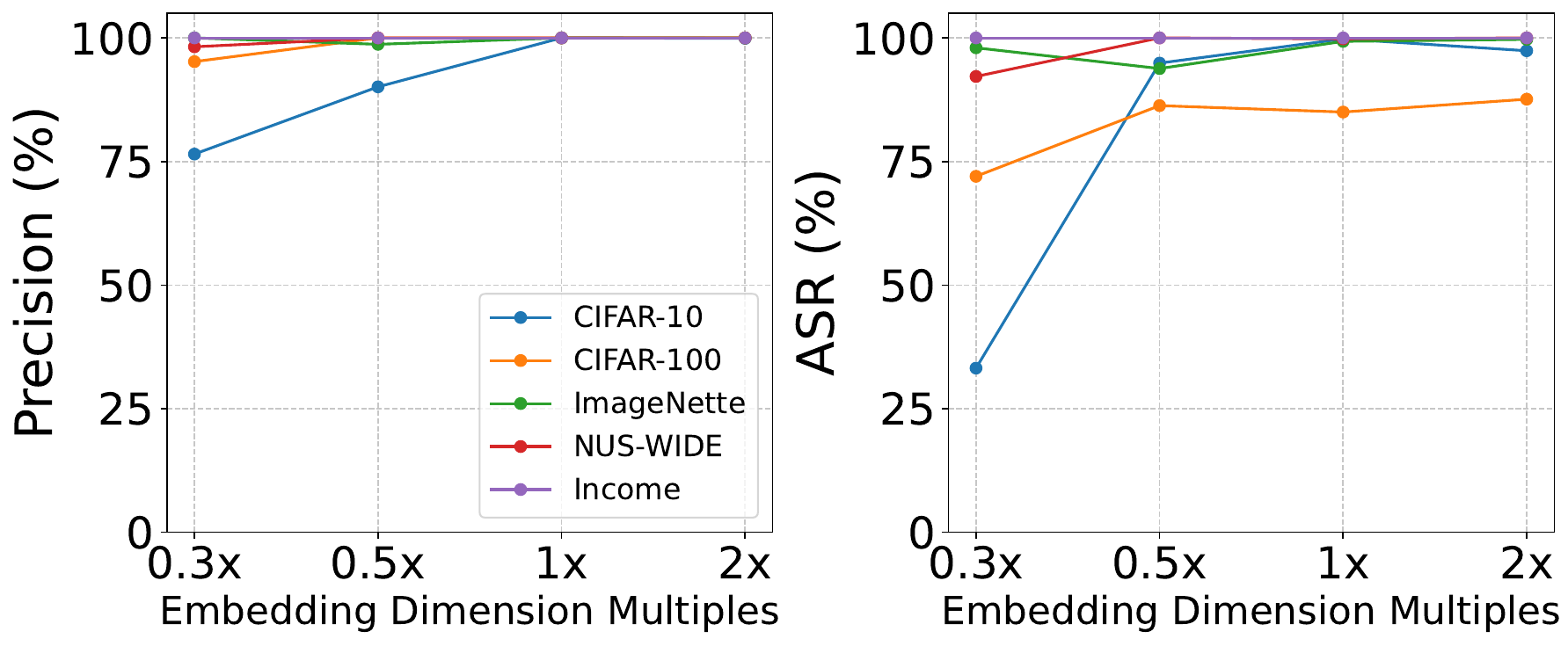}
    \caption{LIA precision and ASR of HASSLE under different sizes of embedding}
    \label{embedding_dimension}
\end{figure}

\subsubsection{Embedding Dimension}
Fig. \ref{embedding_dimension} shows that HASSLE remains largely robust to changes in embedding dimension except when an extremely small embedding dimension is employed for CIFAR-10. The drop in LIA precision may stem from the high feature similarity between certain CIFAR-10 classes (e.g., dog vs. cat, ship vs. airplane), which causes embeddings from different classes to align in similar directions, leading to increased confusion in LIA. The ASR of HASSLE is then affected by both the weaker LIA and a reduced space for adversarial perturbations. However, squeezing the embedding dimension inevitably weakens the MTA as the transmitted information from passive parties is reduced. 

\begin{figure}[t]
    \centering
    \includegraphics[width=0.95\linewidth]{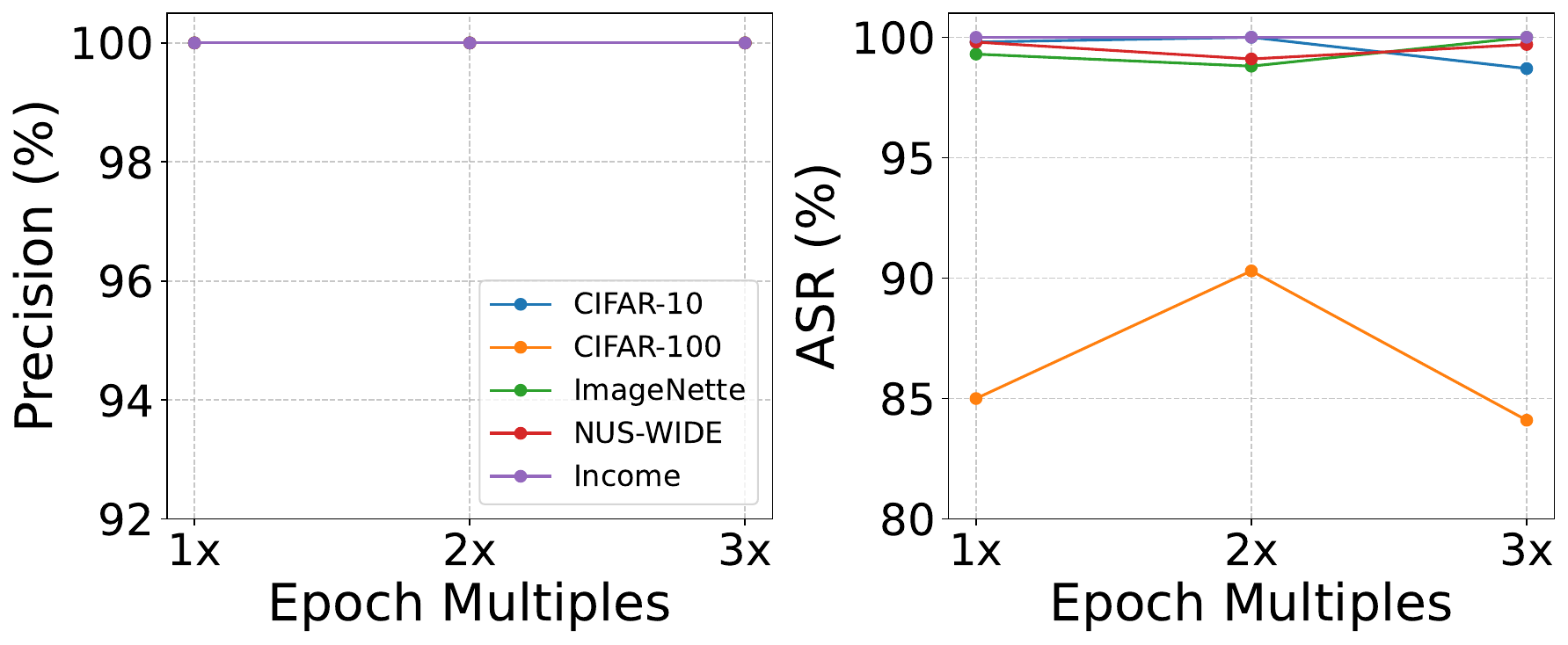}
    \caption{LIA precision and ASR of HASSLE under different total training epochs}
    \label{num_epoch}
\end{figure}

\subsubsection{Number of Epochs}
HASSLE achieves high ASR by exerting dominant influence on the top model from the early stage of training through SSL initialization. To assess HASSLE's attack effectiveness over prolonged training, we extend the number of training epochs to twice and three times the original setting. Fig. \ref{num_epoch} shows that HASSLE maintains the same ASR even under extended training.

\begin{table}[h] 
    \renewcommand{\arraystretch}{1.1}
    \centering
    \caption{Defense Taxonomy}
    \begin{tabular}{|c|c|c|}
        \hline
                                                                           &\textbf{Training-Time}                                                        & \textbf{Post-Training}                                                           \\ \hline
        \textbf{Against LIA}                                                   & \begin{tabular}[c]{@{}c@{}}DPSGD\\ GC\end{tabular} & $\times$                                                            \\ \hline
        \begin{tabular}[c]{@{}c@{}}\textbf{Against}\\ \textbf{Hijacking Attack}\end{tabular} & \begin{tabular}[c]{@{}c@{}}ABL\\ Anomaly Detection\end{tabular} & \begin{tabular}[c]{@{}c@{}}ANP, LIMIT\\ VFLIP, EP\end{tabular} \\ \hline
        \end{tabular}
        \label{defense}
    \end{table}

    \begin{table*}[t]
        \centering
        \setlength{\tabcolsep}{5pt}
        \caption{Robustness Evaluation (results in \%) of HASSLE Against Potential Defenses}
        \begin{tabular}{@{}cc|cccc|ccc|ccc|ccc|c|cc|c@{}}
            \toprule
            \multirow{2}{*}{Dataset} & \multirow{2}{*}{} & \multicolumn{4}{c|}{DPSGD ($\sigma_g$)} & \multicolumn{3}{c|}{GC ($\lambda$)} & \multicolumn{3}{c|}{ABL ($E_{abl}$)} & \multicolumn{3}{c|}{ANP ($N_p$)} & \multirow{2}{*}{VFLIP} & \multicolumn{2}{c|}{EP ($z$)} & \multirow{2}{*}{LIMIT} \\[0.5ex]
            &  & 1e-4 & 5e-4 & 1e-3 & 2e-3 & 0.1 & 0.2 & 0.3 & $\frac{1}{4}E$ & $\frac{1}{3}E$ & $\frac{1}{2}E$ & 10 & 30 & 50 &  & 0.5 & 1 & \\
            \midrule
            \multirow{3}{*}{CIFAR-10}   & LIA & 88.4 & 71.4 & 58.7 & 40.2 & 100 & 100 & 100 & 75   & 67   & 43.9  & 100 & 100 & 100 & 100  & 100  & 100  & 100 \\
                                        & ASR & 63.5 & 61.3 & 55.7 & 47.5 & 100 & 100 & 100 & 96.2 & 64.4 & 27.5  & 93.8 & 89.1 & 4.3   & 38.4 & 92.5 & 69.6 & 73.5 \\
                                        & MTA & 83.3 & 82.9 & 81.3 & 79.9 & 73.3 & 81.6 & 82.4 & 82.3 & 81.8 & 81.8  & 82.8 & 62.5 & 44.2  & 76.5 & 78.1 & 65.3 & 79.3 \\
            \midrule
            \multirow{3}{*}{CIFAR-100}  & LIA & 88   & 40   & 34.4 & 4    & 92.8 & 96   & 99.2 & 64.8 & 59.2 & 48    & 100 & 100 & 100   & 100  & 100  & 100  & 100 \\
                                        & ASR & 57.1 & 31.1 & 41.8 & 0    & 20.7 & 47.7 & 67.7 & 48.7 & 50.8 & 77.6  & 85.3 & 59.4 & 53.3  & 2.5  & 70.6 & 62.3 & 50.6 \\
                                        & MTA & 50.4 & 50.1 & 48.1 & 45   & 45.8 & 49.3 & 51.6 & 51.5 & 51.5 & 50.3  & 41.1 & 22.1 & 18.8  & 47.8 & 50.6 & 47.7 & 49.5 \\
            \midrule
            \multirow{3}{*}{ImageNette} & LIA & 99.6 & 94.5 & 89.8 & 47   & 100  & 100  & 100  & 73.3 & 73.3 & 66.1  & 100 & 100 & 100   & 100  & 100  & 100  & 100 \\
                                        & ASR & 100  & 92.4 & 100  & 62.8 & 100  & 100  & 100  & 84.3 & 79.5 & 85.7  & 74.1 & 4.7  & 0     & 92.5 & 99.4 & 83.2 & 91.7 \\
                                        & MTA & 78.7 & 78.1 & 78.4 & 77.7 & 69.7 & 76.9 & 78.9 & 77.6 & 77.8 & 77.9  & 72.2 & 63.9 & 46.2  & 78.6 & 72.4 & 60.6 & 74.6 \\
            \midrule
            \multirow{3}{*}{NUS-WIDE}   & LIA & 100  & 100  & 87   & 47.3 & 100  & 87.2 & 98.2 & 100  & 100  & 100   & 100 & 100 & 100   & 100  & 100  & 100  & 100 \\
                                        & ASR & 95.4 & 90.3 & 64.4 & 28.8 & 100  & 100  & 100  & 99.4 & 99.3 & 99.4  & 97.2 & 90.5 & 86.2  & 89.1 & 98.8 & 94.5 & 98.7 \\
                                        & MTA & 66   & 65.7 & 66.2 & 66   & 49.6 & 56.3 & 58.2 & 66   & 65.9 & 65.6  & 59.8 & 56.5 & 51.1  & 63.3 & 64.5 & 59.8 & 65.2 \\
            \midrule
            \multirow{3}{*}{Income}     & LIA & 100  & 100  & 100  & 100  & 100  & 100  & 100  & 100  & 100  & 100   & 100 & 100 & 100   & 100  & 100  & 100  & 100 \\
                                        & ASR & 100  & 100  & 100  & 100  & 100  & 100  & 100  & 100  & 100  & 100   & 100 & 100 & 100   & 76.7 & 99.8 & 94.2 & 50.2 \\
                                        & MTA & 64.7 & 64.4 & 65.4 & 67.9 & 65.4 & 65.2 & 64.6 & 65.6 & 65.5 & 65.5  & 65.4 & 60.3 & 50    & 63.7 & 61.1 & 54.8 & 62.7 \\
            \bottomrule
        \end{tabular}
        \vspace{-10pt}
    \end{table*}

\subsection{Robustness Against Possible Defenses}
We systematize the defense against an adversarial attack on VFL model according to time and objective in Table \ref{defense}. The defense can be performed either at training time or post training on either the LIA or hijacking attack module. 

\subsubsection{Differentially-Private SGD (DPSGD)\cite{abadi2016deep}}
This privacy preserving technique is commonly employed in distributed learning scenarios. In our experiment, we adapt the original DPSGD to only clip and perturb the intermediate returned gradients. The gradient norm is clipped to 0.1 for the three RGB image datasets and 0.2 for NUS-WIDE and Income. We evaluate its defense performance against HASSLE under 4 different noise scales $\sigma_g$. It can be observed that a large noise scale considerably reduces our LIA precision since the independent noise at each dimension of the gradient greatly shifts its direction. However, this comes at the cost of lowering model accuracy and the ASR can still be threateningly high unless the noise applied is considerably high.

\subsubsection{Gradient Compression (GC) \cite{Strom2015}}
GC selects a subset of dimension with larger absolute value in the original gradient to update the model parameters. We apply the compression to the returned gradients with three different compression rates $\lambda$. The results demonstrate that GC has limited impact on HASSLE attack even with visible degradation in MTA. Notably, the attack performance is reduced on CIFAR-100 as using only partial dimensions of returned gradient for optimization is not enough to reach a high attack effectiveness. 

\subsubsection{Anti-Backdoor Learning (ABL) \cite{li2021anti}}
ABL first detects suspicious samples with low loss value by regularizing the loss of each training sample above a certain value for $E_\emph{abl}$ consecutive epochs and then unlearns these detected samples in subsequent training. Since the active party has no knowledge about the attack epoch $E_a$ of HASSLE, we perform ABL on the top model using three different settings of $E_\emph{abl}$ around the value of $E_a$, which are 1/2, 1/3, 1/4 of the number of total epochs $E$. Remarkably, the LIA precision is considerably weakened on the three image datasets because ABL's modification to the loss function reverses the gradient direction of those training samples with lower loss value. While the reduced LIA precision affects the ASR of HASSLE, the unlearning process of ABL fails to detect our adversarial embedding that may be injected only for a few epochs before $E_\emph{abl}$ or even after it. Moreover, this embedding is constantly evolving over different epochs, making it difficult to be unlearned.

\subsubsection{Adversarial Neuron Pruning (ANP) \cite{wu2021adversarial}}
ANP selectively prunes sensitive neurons based on a small amount of clean samples to suppress the backdoor attack. We perform ANP on the top model using 1\% of clean bottom model embeddings submitted in the last epoch, after the end of VFL training. Due to the small number of parameters in our top model, pruning any neurons can result in a significant MTA drop. We prune 10, 30, 50 neurons in the top model with the smallest mask value optimized by the ANP algorithm. The results indicate that a large number of neurons have to be pruned to effectively suppress HASSLE, which greatly compromises the MTA.

\subsubsection{VFLIP \cite{cho2024vflip}}
VFLIP is a recently proposed backdoor defense method for VFL. It leverages the submitted embeddings from all parties in the last epoch to train a masked auto-encoder (MAE), which will be used in the test phase to reconstruct submitted embeddings from each individual party. The deviation between the original and reconstructed embeddings is used to detect the anomalies. As illustrated, VFLIP causes a noticeable drop in the ASR of HASSLE with a moderate MTA reduction. However, VFLIP has inconsistent defense performance on NUS-WIDE and ImageNette.
    
\begin{figure}[h]
    \begin{subfigure}{0.48\linewidth}
        \includegraphics[height=3.cm]{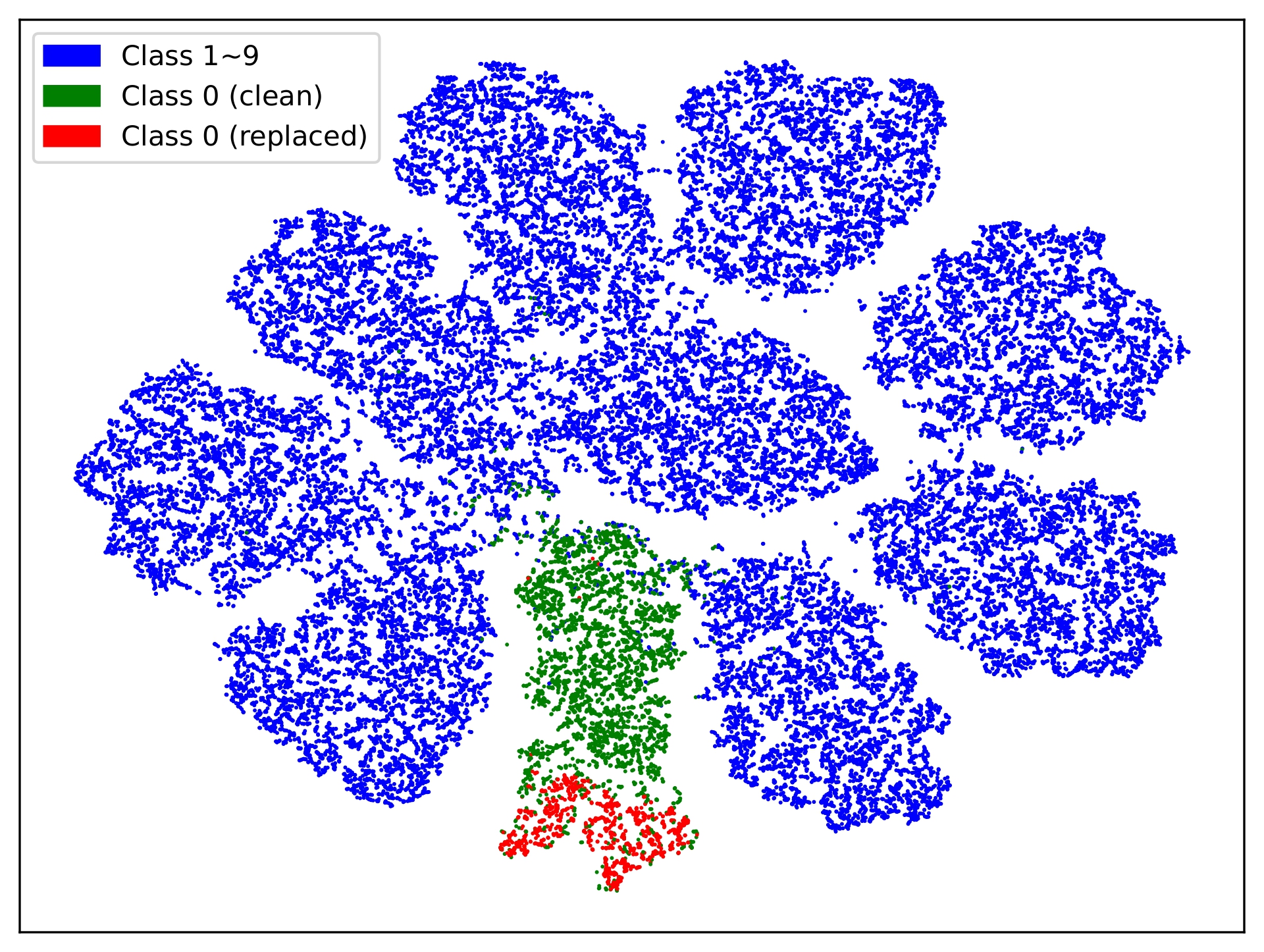}
        \caption{}
        \label{tsne}
    \end{subfigure}
    \begin{subfigure}{0.51\linewidth}
        \includegraphics[height=3.cm]{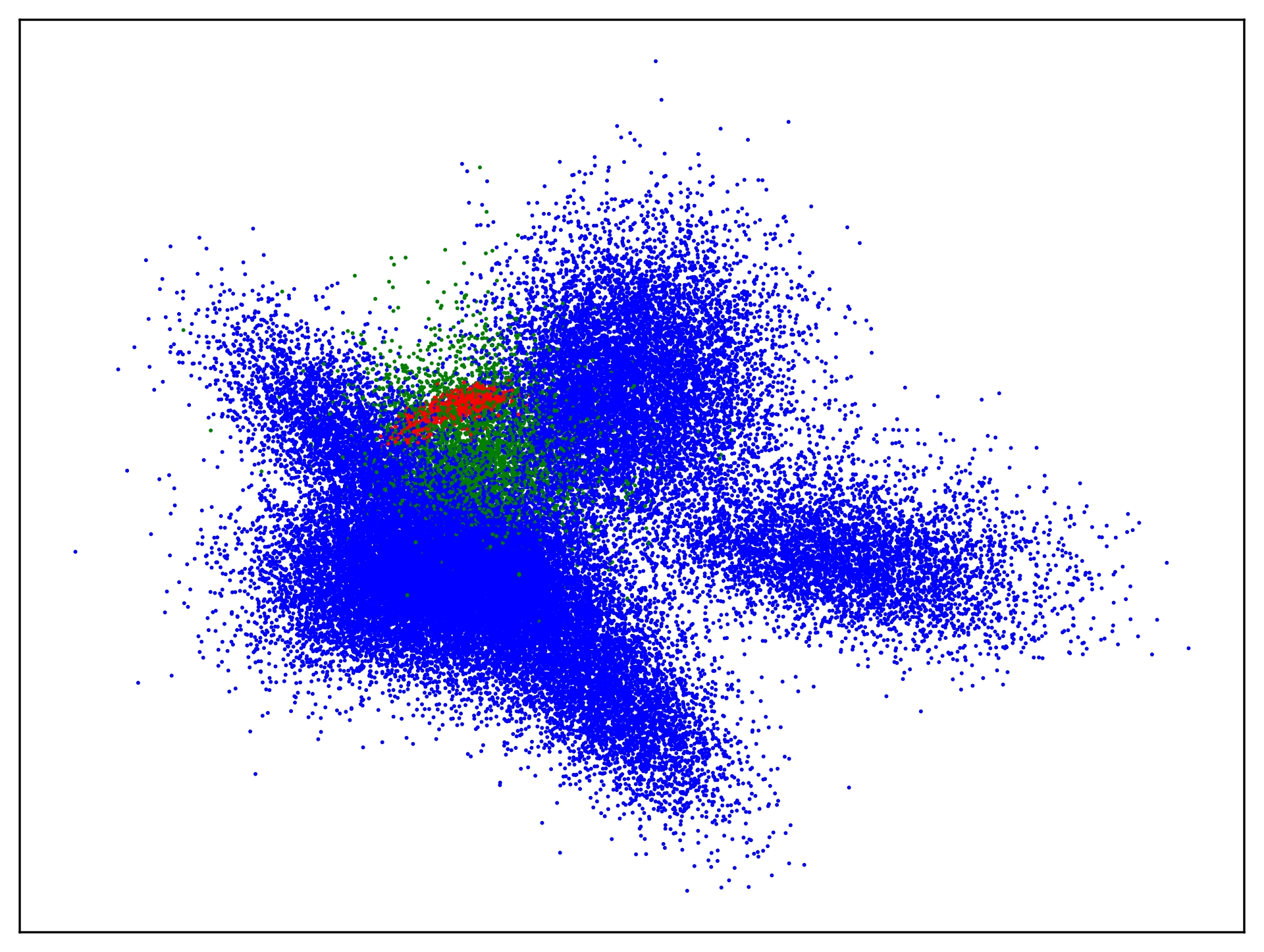}
        \caption{}
        \label{pca}
    \end{subfigure}
    \caption{(a) t-SNE visualization and (b) PCA visualization of embedding dimension submitted by the passive party}
    \label{visualization}
\end{figure}

\begin{table}[b]
    \caption{Anomaly detection using PCA Reconstruction Error and Mahalanobis Distance as key metric.}
    \label{anomaly}
    \setlength{\tabcolsep}{1.6pt}
    \renewcommand{\arraystretch}{1.1}
    \begin{tabular}{@{}c|ccccc|ccccc@{}}
    \toprule
               & \multicolumn{5}{c|}{\textbf{PCA Reconstruction Error}}                                              & \multicolumn{5}{c}{\textbf{Mahalanobis Distance}}                             \\
               & \multicolumn{2}{c|}{Population}    & \multicolumn{3}{c|}{Adversarial} & \multicolumn{2}{c|}{Population}    & \multicolumn{3}{c}{Adversarial} \\
               & mean  & \multicolumn{1}{c|}{std}   & max       & mean      & std      & mean  & \multicolumn{1}{c|}{std}   & max       & mean     & std      \\ \midrule
    CIFAR-10   & 0.089 & \multicolumn{1}{c|}{0.086} & 0.165     & 0.096     & 0.018    & 3.043 & \multicolumn{1}{c|}{0.862} & 3.109     & 2.83     & 0.081    \\
    CIFAR-100  & 0.479 & \multicolumn{1}{c|}{0.286} & 0.63      & 0.444     & 0.058    & 9.844 & \multicolumn{1}{c|}{1.761} & 9.648      & 8.162    & 0.305    \\
    ImageNette & 0.103 & \multicolumn{1}{c|}{0.117} & 0.093     & 0.029     & 0.01     & 3.028 & \multicolumn{1}{c|}{0.911} & 3.383     & 2.527    & 0.107    \\
    NUS-WIDE   & 0.079 & \multicolumn{1}{c|}{0.095} & 0.108     & 0.061     & 0.018    & 2.974 & \multicolumn{1}{c|}{1.075} & 4.192     & 3.618    & 0.099    \\
    Income     & 0.116 & \multicolumn{1}{c|}{0.168} & 0.069     & 0.027     & 0.003    & 2.904 & \multicolumn{1}{c|}{1.251} & 3.712     & 3.127    & 0.104    \\ \bottomrule
    \end{tabular}
    \end{table}

The following presents the adaptive defenses for our attack.

\subsubsection{Anomaly Detection}
Fig. \ref{visualization} illustrates the relative position of adversarial embeddings within the overall embedding distribution during the first attack epoch on the CIFAR-10 dataset, after applying two dimensionality reduction techniques. These embeddings typically reside within the main cluster, making them difficult to separate. To quantify their stealthiness, we use two anomaly detection methods. The first employs PCA to compress embeddings and filters outliers based on reconstruction error. The second measures the Mahalanobis distance between each embedding and the overall distribution. We focus on the first attack epoch, where adversarial embeddings initially appear and deviate most from the normal distribution. As shown in Table \ref{anomaly}, the maximum error/distance for HASSLE adversarial embeddings remains below the population mean plus one standard deviation, highlighting their stealthiness, which makes anomaly detection challenging without incurring a high false positive rate.

\subsubsection{Embedding Perturbation (EP)}
Since HASSLE necessitates a well-optimized adversarial embedding to achieve high ASR, we adopt similar idea from Randomized Smoothing \cite{cohen2019certified} to design a defense through embedding perturbation. In test time, EP adds random Gaussian noise to the submitted embeddings from the adversary for 100 different times, and obtain the prediction with the most occurrences. The parameter $z$ denotes the multiples of noise standard deviation to that of the training samples. The result shows that EP has considerable impact on HASSLE ASR in CIFAR-10 and CIFAR-100 when setting $z=1$, at the cost of noticeable MTA drop.

\subsubsection{Influence Limitation (LIMIT)}
By design, the top model output is directly affected by the magnitudes of the submitted embeddings and the corresponding layer weights. To mitigate the adversary's impact, we constrain the norm of its submitted embeddings to match that of the active party and limit the norm of the Fully-Connected layer weights associated with the adversary to be equal to those of the active party. The result shows that this heuristic significantly affects HASSLE’s ASR in the CIFAR-100 and Income datasets, and noticeably reducing its ASR on CIFAR-10. However, the MTA suffers mild degradation because the classification task may inherently rely more on the features provided by the adversary.

In summary, there is no one-size-fits-all defense against HASSLE. However, all defenses exhibit effectiveness in reducing the ASR of HASSLE, with DPSGD, ANP and VFLIP having the highest impact. In terms of MTA preservation, DPSGD, GC and ABL affect main task the least. Taking into account the computation overhead,  the additional computation of DPSGD, GC, ABL and LIMIT are negligible since they only perform noise addition, sorting and norm scaling. Others incur high computation cost. While ANP and VFLIP are performed after VFL training, EP is conducted by forwarding the top model multiple times during each inference, which significantly reduces  the inference speed. Considering the trade-off among the ASR reduction, main task accuracy and computation overhead, we identify DPSGD, VFLIP and LIMIT as suitable defenses against HASSLE. Since HASSLE can still potentially pose a serious threat to VFL system deployed with recommended defenses, we envisage a defense-in-depth strategy that considers both LIA and hijacking attacks. Specifically, the active party can employ a loss function that reduces the correlation between its returned gradient and label information, and utilize an adversarial feature purification technique like VFLIP during inference.

\vspace{-0.3cm}

\section{Conclusion}
In this work, we propose a novel hijacking attack framework HASSLE to compromise the VFL system. Our attack outperforms prior works due to its more accurate label inference algorithm and more effective adversarial embedding with its influence on the top model enhanced by the SSL applied on the adversary's bottom model. We evaluate HASSLE rigorously to analyze its individual pipeline and combined performance across various settings and against known defenses. These analyses offer valuable insights into the limitations of existing defense strategies in improving the trustworthiness of VFL. 

\bibliographystyle{IEEEtran} 
\bibliography{bibfile} 

\end{document}